\documentstyle[preprint,aps,epsf]{revtex}
\begin{document}
\def\be{\begin{equation}}
\def\ee{\end{equation}}
\def\bea{\begin{eqnarray}}
\def\eea{\end{eqnarray}}
\def\E{{\rm e}}
\def\bearst{\begin{eqnarray*}}
\def\eearst{\end{eqnarray*}}
\def\peleven{\parbox{14cm}}
\def\peffec{\peight{\bearst\eearst}\hfill\peleven}
\def\pspace{\peight{\bearst\eearst}\hfill}
\def\ptwelve{\parbox{15cm}}
\def\peight{\parbox{8mm}}
\title
{Statistical Theory for the Kardar--Parisi--Zhang Equation 
in $1+1$ dimension}
\author
{A. A. Masoudi $^{a,e}$, F. Shahbazi $^{a,e}$, J. Davoudi $^{c,e}$ \\
and M. Reza Rahimi Tabar $^{b,d,e}$}
\address
{\it $^a$ Dept. of Physics , Sharif University of Technology,
P.O.Box 11365-9161, Tehran, Iran.\\
$^b$ CNRS UMR 6529, Observatoire de la C$\hat o$te d'Azur,
BP 4229, 06304 Nice Cedex 4, France,\\
$^c$ Max-Planck Institute for Complex Systems, Noethnitzer St.38, 01187 
Dresden, Germany \\
$^d$ Dept. of Physics , Iran  University of Science and Technology,
Narmak, Tehran 16844, Iran.\\
$^e$ Institute for Studies in Theoretical Physics and 
Mathematics
Tehran P.O.Box: 19395-5531, Iran,\\
}  
\maketitle
\begin{abstract}
The Kardar-Parisi-Zhang (KPZ) 
equation in 1$+$1 dimension dynamically develops sharply connected valley structures
within which the height derivative {\it is not} continuous. 
There are two different regimes 
before and after creation of the sharp valleys.
We develop a statistical theory for the KPZ equation
in 1+1 dimension driven with a random forcing which is white in time and 
Gaussian correlated in space.
A master equation is derived for the joint 
probability density function of height difference and height gradient
$P(h-\bar h,\partial_{x}h,t)$ when the forcing correlation length is much 
smaller than the system size and much bigger than the typical sharp valley width. 
In the time scales before the creation of the sharp valleys
we find the exact generating function of $h-\bar h$ and $\partial_x h$.
Then we express the time scale when the sharp valleys develop, in terms of 
the forcing characteristics. 
In the stationary state, when the sharp valleys are fully developed, 
finite size corrections to the scaling laws of the structure functions  
$ \langle (h-\bar h)^n (\partial_x h)^m \rangle$ are also obtained.
\end{abstract}
\newpage

\section{Introduction}
There has been a great deal of recent work on the formation, growth and 
geometry of interfaces [1-5]. The dynamics of interfaces has turned out to be
one of the most fascinating and at the same time challenging topics in 
theoretical non-equilibrium physics. There has been two principal approaches 
for theoretical analysis of such problems. The first is based on computer 
simulations of discrete models and often provides useful links between 
analytic theory and
experiments. The second approach aims to describe the dynamical process by 
a stochastic differential equation. This procedure neglects the short 
length-scale details but provides a coarse-grained description of the 
interface (that is suitable for characterising the asymptotic 
scaling behaviour). Theoretical modelling of growth processes started with the 
work by Edwards and Wilkinson (EW) [6]. They suggested that one might describe 
the dynamics of the height fluctuations by a simple linear stochastic equation.
Kardar--Parisi--Zhang  (KPZ) [7], realized that there is a relevant term 
proportional to the square of the height gradient which represents a 
correction for lateral growth. Indeed the KPZ equation is a prototype 
model for
a system in which the interface growth is subjected to a random external 
flux of particles.
The randomness is described by an annealed random noise which mimics the 
random adsorption of molecules onto a surface. In the KPZ model 
(e.g in the 1+1 dimension), the surface 
height $h(x,t)$ on the top of location $x$ of 1-dimensional substrate
satisfies a stochastic random equation,

\be{\label {kpz}}
\frac{\partial h}{\partial t} - \frac{\alpha}{2}(\partial_x h)^2 = \nu \partial^
2_x h+f(x,t),
\ee

where $\alpha \geq 0 $ and $f$ is a zero-mean, statistically homogeneous, white in time and Gaussian  
process with covariance
\be{\label {fc}}
\langle f(x,t)f(x',t')\rangle =2 D_0 \delta(t-t') D(x-x'),
\ee
where,
\be{\label {fv}}
D(x-x') = \frac{1}{\sqrt{\pi} \sigma} \exp(- \frac{ (x-x')^2}{\sigma^2}), 
\ee
and $\sigma$ is the variance of $D(x-x')$. 
 Typically the correlation of forcing  
is considered as delta function for mimicking the short range correlation. 
We regularise the delta function correlation by a Gaussian function.
When the variance $\sigma$ is much less than the system size we would expect that  
the model would represent a short-range correlated forcing. So we would
stress that our calculations are done for finite $\sigma\ll L$, where $L$ is the  
system size. 
The average force on the interface is unimportant and may be removed from the 
equation of motion. Every term in the eq.(\ref{kpz}) involves a specific 
physical 
phenomenon contributing to the surface evolution. The parameters $\nu$, 
$\alpha$  and $D_0$ ( and $\sigma$) are describing the surface 
diffusive relaxation, non-linear lateral growth and the effective noise
strength, respectively. 

We consider a substrate of size $L$ and define the mean height of growing 
film and its roughness $w$ by,
\be{\label {hm}}
\bar{h} (L,t) = \frac{1}{L} \int_{-L/2} ^{L/2} dx h(x,t),
\ee

\be{\label {wid}}
w(L,t) = ( \langle ( h - \bar{h})^2 \rangle )^{1/2},
\ee

where $\langle \cdots \rangle$ denotes an averaging over different 
realizations of the noise (samples). 
Starting from a flat interface (one of the possible initial conditions),
it was conjectured by Family and Vicsek [8] that a scaling of space by factor
$b$ and of time by a factor $b^z$ ($z$ is the dynamical scaling exponent), re-scales 
the roughness $w$ by factor $b^{\chi}$ as follows,
\be
w( bL, b^zt) = b^{\chi} w( L,t),
\ee
which implies that
\be
w(L,t) = L^{\chi} f(\frac{t}{L^z}).
\ee
If for large $t$ and fixed $L$ $(\frac{t}{L^z} \rightarrow \infty)$, $w$ 
saturates then $f(x) \rightarrow  const.$  as  $ x \rightarrow \infty$.
However, for fixed large $L$ and $1<< t << L^z$, one expects that 
correlations of the height fluctuation are set up only within a distance 
$t^{1/z}$  and thus must be independent of $L$. This implies that for $x << 1$,
$f(x)\sim x^{\beta}$ with $\beta=\frac{\chi}{z}$. Thus dynamic scaling 
postulates that [9],
\bea
w(L,t) &&\sim t^{\beta} \hskip 2cm  1 << t << L^z  \cr \nonumber \\
       &&\sim L^{\chi}  \hskip 2cm   t >> L^z
\eea
The roughness exponent $\chi$ and the dynamic exponent $z$ characterise the 
self-affine geometry of the surface and its dynamics, respectively. 
Several time-regimes can be distinguished in the time evolution of the 
surface roughness. They can be summarised as follows:
for very early times , the noise term dominates since its contribution to 
the equation grows as the square root of time. In this time-regime, 
the surface roughness grows as $w(t)\sim t^{1/2}$. For intermediate times, 
the linear term has the main contribution. The linear case ($\alpha=0$)  is the 
Edwards-Wilkinson model for which one can easily find that the surface 
roughness 
behaves as $w(t)\sim t^{\beta_0}$, where the value of $\beta_0$ depends on 
dimension of substrate($\beta_0=\frac{2-d}{4}$ 
for d-dimensional surface).
For later times, the contribution of the relevant non-linear term 
becomes a
dominant one and the surface roughness growth is characterised by the 
behaviour $w(t)\sim t^{\beta}$ . 
For very late times and finite substrate-length $L$, 
the roughness saturates to the value $w(t\rightarrow \infty,L)\sim L^{\chi}$.
Of course, in an experiment or in a numerical simulation the transition 
between the different regimes is not sharp and different crossover 
behaviours can be observed.
Galilean invariance implies the relation $\chi + z = 2$ independent of 
dimension [10-11]. It means that there is only one independent exponent  
in the KPZ dynamics. 
In the one-dimensional substrates a fluctuation-dissipation 
theorem yields exactly $z=\frac{3}{2}$, $\chi=\frac{1}{2}$, 
$\beta=\frac{1}{3}$ [12].
In contrast to 1-dimension, the case $d\geq 2$ can be only attacked 
by approximative field-theoretic perturbative expansions [13-16]. 
It is well-known that 
 effective coupling constant for the KPZ equation is 
$ g = \frac{2 \alpha^2 D_0}{\nu^3}$. 
Phase diagram information extracted from the renormalisation group flow 
indicates that $d=2$ plays the role of a lower critical dimension. 
For $d \leq 2$, the Gaussian fixed point ($\alpha = 0) $ is infrared-unstable,
and there is a crossover to the stable strong coupling fixed point. For $d >2$,
a third fixed point exists, which represents the roughening transition. It is 
unstable and appears between the Gaussian and strong coupling fixed points
which are now both stable. 
Only the critical indices of the 
strong-coupling regime ($g \rightarrow \infty $ or $\nu \rightarrow 0$)
are known in 1+1 dimensions and their 
values in higher dimensions as well as 
properties of the roughening transition have been known only numerically 
[17-23], and the various approximation schemes [24-32].

The theoretical richness of the KPZ model is partly due to close 
relationships with other areas of statistical physics. It is shown  
that there is a mapping between the equilibrium statistical mechanics of 
a two dimensional smectic-A liquid crystal onto the non-equilibrium dynamics
of the (1+1)- dimensional stochastic KPZ equation [33]. It has been shown 
in [34] that, one can map the kinetics of the annihilation process $A + B \rightarrow 0$
with driven diffusion onto the (1+1)-dimensional KPZ equation. Also  
the KPZ equation is closely related to the dynamics of a sine-Gordon chain 
[35],
the driven-diffusion equation [36-37], high $T_c$- superconductor [38] 
and directed paths in the random media 
[39-52]
and  charge density waves [53], dislocations in disordered solids [3],
formation of large-scale structure in the universe [54-57] , Burgers
turbulence [58-85,90] and etc.

As mentioned the main difficulty with the KPZ equation is that it is 
controlled, in all dimensions, by a strong disorder ( or strong coupling)
fixed point and efficient tools are missing to calculate the exponents and 
other universal properties e.g. scaling functions, amplitudes, etc. 
Despite the fact that in one dimension, the exponents are known, but many
properties, including the probability density function (PDF) of the height
of a growing interface have been so far measured only in numerical 
simulations.
Recently, it is shown that for one particular model of the KPZ class, 
the asymmetric exclusion process (ASEP), the whole distribution of displacement
of particles could be calculated for a finite geometry by Bethe ansatz [86-87].
As discussed in [86-87], the PDF of displacement of particles is related to
the PDF of height differences in the growth process. It is proved in [86-87]
that the PDF of height differences has the following asymptotic:
\bea
P(\frac{h-\bar h}{w}) \hskip 1.5 cm && \sim  \hskip 2cm \exp(-A y^{5/2})   
\hskip 1cm  y \rightarrow +\infty, \cr \nonumber \\
&&\sim \hskip 2 cm \exp(-B |y|^{3/2}) \hskip 1cm  y \rightarrow -\infty,
\eea
where $y= \frac{h-\bar h}{w}$ and $w$ is the variance of the height fluctuation
about its average. This result is in agreement with the numerical 
experiments [88]. 

In this paper we are interested in the statistical 
properties of the KPZ equation in the strong coupling limit 
($\nu \rightarrow 0$). 
The limit is singular,i.e. through which the surface develops sharp valleys. So 
starting with a flat surface after a finite time scale, $t_c$, 
the sharp valley singularities are dynamically developed. In the singular points 
(sharp valleys) spatial derivative of the $h(x,t)$ is not 
continuous. Hence the limit of $\nu\rightarrow 0$ is not singular 
for $t<t_c$, and we can ignore the diffusion term while after developing 
the singularities the diffusion term has finite contribution in the 
PDF of height fluctuations. Inspired by the methods proposed, recently in 
the works done by Weinan E and Vanden Eijnden [73],
we develop a statistical method to describe the 
moments of height difference and height gradient of height field $h(x,t)$.
We derive a master equation
for joint probability density function (PDF) of the height difference
$(h - \bar h)$ and height gradient $\partial_x h$, 
$P( ( h - \bar h ), \partial_x h )$ for given $g$ or diffusion constant 
$\nu$. 
We will consider two different time scales in the limit of $\nu \rightarrow
0$. ($i$) Early stages before developing the 
sharp valley singularities and ($ii$) established stationary state comprising 
with fully developed sharp valley singularities. In 
the regime ($i$) 
ignoring the relaxation term in the equation of the joint PDF when $ \nu 
\rightarrow 0$, we determine the exact generating function of joint 
moments of height and height gradient fields. Realisability condition 
for the resulting joint PDF sheds light on the time scale of the 
sharp valley formation. 
In contrary the limit $\nu \rightarrow 0$ is singular 
in the regime ($ii$) leading to an unclosed term (relaxation term) 
in the PDF equation. However we show that the unclosed term 
can be expressed in terms of 
statistics of some quantities defined on the singularities 
(sharp valleys).
Identifying each sharp valley in position $y_0$ with three quantities, namely 
the gradient of $h$ in the position $y_{0^+}$, $ y_{0^-}$ and its height
from the $\bar h$, we determine the dynamics of these quantities. 
In the both regimes all the moments,  
$\langle (h-\bar h)^n (\partial_x h)^m \rangle $, for given 
$n$ and $m$ are found.
In the regime ($ii$) we will prove that in leading order, when 
$L\rightarrow \infty$, fluctuation of the height field is not 
intermittent and succeed to give the analytic form of  
amplitudes of the all structure functions. Besides, the scaling 
behaviour and the amplitudes of all the correction 
terms due to the finite size effect are calculated.

The paper is organised as follows; in section two, we derive the master
equation for the joint PDF of height differences and height gradients
for given diffusion $\nu$. 
We convert the height PDF,i.e. $P(h-\bar h,t)$, evolution equation 
consequently to a Fokker-Planck 
equation for an arbitrary given diffusion constant.
In section three we will consider the limit of $\nu \rightarrow 0$ of the 
master equation in the time scales that there are no any singularities in 
the 
surface (before developing the sharp valleys). We determine the exact and explicit 
expression of the generating function for the moments 
$ \langle (h - \bar h)^n (\partial_x h)^m \rangle$ for given $n$ and $m$.
In section 4 we consider the master equation in the limit $\nu \rightarrow 0$
and consequently when the singularity are fully developed. 
In this regime the relaxation term has finite contribution in 
the master equation.
Using the methods introduced in [73], we prove that the unclosed term can be
written in terms of quantities which are defined on sharp valleys where
$\partial_x h$ is discontinuous. Also in this section we determine the 
relation  between the density of sharp valleys and the forcing variance 
$k_{xx}(0)$ in detail. 
In section 5, we derive the moments of height fluctuation in the stationary
state and show that the PDF of $(h -\bar h)$ is strongly asymmetric and prove
that in to leading order the n-th moments of $(h-\bar h)$, 
i.e. $\langle (h - \bar h)^n \rangle$, 
can be written in terms of the second order moment of height fluctuation
in a non-intermittent way. 
We determine the amplitudes of the all of moments and show that  
the amplitudes of moments $\langle (h - \bar h)^n (\partial_x h)^m
\rangle$ can be written in terms of characteristics of singularities. We also
derive the finite size effect on the moments of height differences and
determine the amplitudes of all correction terms. In section 6, we
derive the PDF of quantities which characterise the singularity, and 
therefore find the equation for their evolution. 

\section{ The master equation for height difference and height 
gradient }
In this section we consider the $1+1-$dimensional KPZ equation and derive  
the master equation to describe the joint-PDF of height-difference and
height-gradient i.e. $P(h-\bar{h},\partial_{x} h)$, for given $\nu$ and 
$\alpha$. It is shown that the equation for the joint PDF is not closed  
due to the $linear$ term $\nu\partial^{2} h$. The PDF of height 
difference is related to the joint PDF, $P(h-\bar{h},\partial_{x} h,t)$  
by the relation 
$P(h-\bar{h},t)=\int_{- \infty} ^{\infty} P(h-\bar{h},
\partial_{x} h,t)d(\partial_{x} h) $.
We show that $P(h-\bar{h},t)$ satisfies a Fokker-Planck equation and 
write down the explicit expression of drift and diffusion coefficient 
$D^{(1)}$ and $D^{(2)}$. It is shown that the drift and the diffusion 
coefficients can be written  
in terms of the conditional average of 
$\langle (\partial_{x} h)^{2}|h-\bar h\rangle$.

We consider a one-dimensional line of length $L$ and a surface of
height $h(x,t)$, and gradient $ \partial_x h(x,t)$ at time $t$. 
The 1+1 dimensional 
KPZ equation governed over height field $h(x,t)$ is defined in 
eq.(\ref{kpz}) while $u(x,t)=-\partial_x 
h(x,t)$ is a solution of the the so called Burgers equation as,
\be{\label{bur}}
u_{t} + \alpha u u_x = \nu u_{xx} - f_x(x,t),
\ee
where the covariance of $f$ is given by eqs.(\ref{fc}) and (\ref{fv}). To 
investigate the 
statistical properties of eqs.(\ref{bur}) and (\ref{kpz}), let us define the 
generating function $Z(\lambda, \mu,x,t)$ as:
\be{\label{gen-i}}
 Z(\lambda,\mu,x,t) = \langle\exp(-i\lambda (h(x,t)-\bar h)-i\mu u(x,t))\rangle.
\ee
It follows from eqs.(\ref{bur}) and (\ref{kpz}) that generating function 
$Z$ is a solution of the following equation,

\bea{\label{gen-eq1}}
Z_t&&=i\gamma(t)\lambda Z-i\lambda\frac{\alpha}{2} Z_{\mu\mu}-\lambda^{2}k(0)Z
+\frac{i\alpha\lambda+\nu\lambda^{2}}{\mu}Z_{\mu} 
-\frac{\nu\lambda}{\mu}Z_x \cr \nonumber \\
&&-i\alpha \mu(\frac{Z_x}{\mu})_{\mu}
+\mu^2k_{xx}(0)Z-i\mu\nu\langle u_{xx}\exp(-i\lambda (h-\bar h)-i\mu 
u)\rangle,
\eea

where $k(x-x')=2D_{0}D(x-x'),\gamma(t)={\bar h}_t$, 
$k(0) = \frac{D_0}{\sqrt{\pi} \sigma}$ and 
$ k_{xx}(0) = -\frac{2D_0}{\sqrt{\pi} \sigma^3}$. To derive 
eq.(\ref{gen-eq1}) we have used the following identities:
\be
\langle h_{xx} \exp{  (-i\lambda(h(x,t)-\bar h)-i\mu u(x,t))} \rangle =
\frac{-i}{\mu} \{ Z_x + \lambda\partial_\mu Z \},
\ee
\be
\langle f(x,t) \exp{  (-i\lambda(h(x,t)-\bar h)-i\mu u(x,t))}\rangle =
-i\lambda k(0) Z,
\ee
and
\be
\langle f_x( x,t) \exp{  (-i\lambda(h(x,t)-\bar h)-i\mu u(x,t))}\rangle =
-i\mu k_{xx}(0) Z,
\ee
where we have used the fact that $D_x(0) =0$.
This is evident that due to the last term in the eq.(\ref{gen-eq1}), it is 
not closed. Assuming statistical homogeneity ($Z_x=0$) we have,
\bea{\label{gen-eq2}}
-i\mu Z_t&&=\gamma\lambda\mu Z-\frac{\alpha}{2}\lambda\mu Z_{\mu\mu}
+i\lambda^2\mu k(0)Z-i\mu^3k_{xx}(0)Z \cr \nonumber \\
&&-i(\nu\lambda^2+i\alpha\lambda)Z_{\mu}
-\mu^2\nu\langle u_{xx}\exp(-i\lambda \tilde{h}(x,t)-i\mu u(x,t))\rangle,
\eea
where ${\tilde h}(x,t)=h(x,t)-\bar h$.
Defining $P({\tilde h},u,t)$ as the joint probability density function of 
$\tilde h$ and $u$, one can construct $P({\tilde h},u,t)$ in terms of 
generating function $Z$ as,
\be{\label{def-pdf}}
P({\tilde h},u,t)=\int\int\frac{d\lambda}{2\pi}\frac{d\mu}{2\pi}
\exp(i\lambda {\tilde h}+i\mu u)Z(\lambda,\mu,t).
\ee
It follows from eqs.(\ref{def-pdf}) and (\ref{gen-eq2}) that $P({\tilde 
h},u,t)$ satisfies the following equation,
\bea{\label{pdf-eq1}}
-P_{ut}&& = -\gamma P_{{\tilde h}u}-\frac{\alpha}{2}(u^2P)_{{\tilde h},u}
-\alpha(uP)_{\tilde h}-k(0)P_{{\tilde h}{\tilde h}u}+k_{xx}(0)P_{uuu}
- \nu (uP)_{\tilde h \tilde h} \cr \nonumber \\
&&-\nu\int\int\frac{d\lambda}{2\pi}\frac{d\mu}{2\pi}
\exp(i\lambda{\tilde h}+i\mu u)\mu^2 \langle u_{xx}\exp(-i\lambda {\tilde h}(x,t)
-i\mu u(x,t))\rangle.
\eea
Now we can rewrite the last term in eq.(\ref{pdf-eq1}) as following,
\bea{\label{anom0}}
&&\nu\int\int\frac{d\lambda}{2\pi}\frac{d\mu}{2\pi}\exp(i\lambda {\tilde h}+i\mu u)      
\mu^2 \langle u_{xx}(x,t)\exp(-i\lambda {\tilde h}(x,t)-i\mu u(x,t))\rangle \cr \nonumber \\
&&=-\nu \langle u_{xx}(x,t)\delta(\tilde{h}(x,t) - \tilde h)\delta(u(x,t)-u)\rangle \cr \nonumber\\ 
&&=-\nu \{\langle u_{xx}|u,{\tilde h}\rangle
P(u,{\tilde h},t)\}_{uu}.
\eea
where $\langle\cdot |u,\tilde h\rangle$ denotes the average 
conditional on a given $u, \tilde h$.
Therefore using the eq.(\ref{anom0}) it follows that the $ P({\tilde 
h},u,t)$ satisfies the following,
\bea{\label{pdf-eq2}}
-P_{ut}&&=-\gamma P_{{\tilde h}u}-\frac{\alpha}{2}(u^2P)_{{\tilde h},u}
-\alpha(uP)_{\tilde h}-k(0)P_{{\tilde h}{\tilde h}u}+k_{xx}(0)P_{uuu}
-  \nu (uP)_{\tilde h \tilde h}+ \cr \nonumber \\
&&+\nu \{\langle u_{xx}|u,{\tilde h}\rangle
P(u,{\tilde h},t)\}_{uu}.
\eea
This equation is exact for a given $\nu$
and as it is clear the fingerprints of diffusion term is giving rise 
again to an unclosed equation for $ P({\tilde h},u,t)$.
Accessing to the functional form of the 
conditional averaging $  \langle u_{xx}|u,{\tilde h}\rangle$ is one of 
the major difficulties in the formulation. 
From eq.(\ref{pdf-eq1}) we see that $P(u,\tilde h)=P(-u,\tilde h)$ which 
results to, 

\be{\label{iden1}}
\langle u | {\tilde h} \rangle = 0,
\ee

for any given $\nu$. In fact for a given $h$ the average of 
height gradient, $u$, is consequently zero.
The identity proposed by eq.(\ref{iden1}) is not
restricted to any limiting asymptotic and is true in all regimes of the 
dynamical evolution of the surface. 
Also eq.(\ref{pdf-eq2}) allows us to determine a dynamical equation 
for $P(h-\bar h)$. 
Doing so, we multiply the eq.(\ref{pdf-eq2}) to $u$ an integrate over $u$ 
from  $ - \infty$ to $ + \infty$, from which we get,
\bea 
\partial_t P(\tilde h, t ) =&&  \frac{\partial}{\partial \tilde h } \{
( \frac{\alpha}{2} ( \langle u^2\rangle - \langle u^2|\tilde h \rangle )) P(\tilde h ,t) \} \cr \nonumber \\
  && + \frac{\partial^2}{\partial \tilde h ^2} \{ (k(0) - \nu \langle u^2|\tilde h\rangle)
  P(\tilde h,t) \},
\eea
where $\tilde h = h - \bar h$ and the relation $\gamma= 
\frac{\alpha}{2} \langle u^2\rangle$ is used. 
This is a Fokker-Planck (FP) equation, describing
the time evolution of $ P(\tilde h,t) $. The drift coefficient in the FP 
equation is, 
\be
D^{(1)} = - \frac{\alpha}{2} 
(\langle u^2\rangle - \langle u^2|\tilde h\rangle ),
\ee
and the diffusion coefficient reads, 
\be
D^{(2)} = k(0) - \nu \langle u^2 | \tilde h \rangle.
\ee
It is evident that aiming to obtain $P(h - \tilde h)$, one should know the 
conditional average $\langle u^2 | \tilde h\rangle$. 
The equation has the following stationary solution:
\be
P_{stat.}(\tilde h) = \frac{N}{D^{(2)}} \exp \{ \int_{\tilde h_0} ^{\tilde h}
d \tilde h' D^{(1)}(\tilde h') / D^{(2)}(\tilde h') \}
\ee

Where $N$ is the normalisation coefficient. 
Therefore to derive the moments of hight difference $h-\bar h$ i.e.
$ \langle (h-\bar h)^n \rangle$, we need the conditional averaging 
$\langle u^2 | \tilde h\rangle$.
The simplified picture given by this equation is indicating that
all the knowledge one would know in order to obtain the behaviour of PDF
is buried in the functional form of one conditional average
, i.e. $ \langle u^2| {\tilde h}\rangle$.
Although simple but it is clear that the conditional average
$ \langle u^2| {\tilde h}\rangle$
would have 
a non-trivial dependence on $\nu$ and $L$ in the limit of $\nu\rightarrow 0$.
Instead of following this strategy however in the next section we follow 
another direct way of extracting the moments of height 
difference $(h-{\bar h})$ in the strong coupling limit ,
i.e. $\nu \rightarrow 0$.

\section{ The joint correlations of height difference and height gradient 
before sharp valley formation}

When $\sigma$ is finite,
the very existence of the non-linear term 
in the KPZ equation leads to the
development of the sharp valley singularities in a 
{\it finite time} and in the strong coupling 
limit($\nu \rightarrow 0$). In one dimension system is already in the 
strong coupling regime so starting from any finite value of $\nu$ in 
large time system develops sharp valley singularities, Fig(2). Therefore one 
would distinguish between 
different time regimes before and after the sharp valley formation.
Starting from a flat initial condition, i.e. $h(x,0)=0, u(x,0)=0$, 
which its evolution is given by inviscid KPZ equation, we know that after a 
finite time the  
derivative of function $h(x,t)$ becomes singular. After this time scale the 
diffusion term is important, but we can neglect that before  
appearance of the singularities. So the equation governing the the evolution 
of generating function, $Z(\mu,\lambda,t)$ before the creation of the sharp valleys
is given by,
\begin{equation}{\label{gen-eq3}}
Z_t=i\gamma(t)\lambda Z-i\lambda\frac{\alpha}{2} Z_{\mu\mu}-
\lambda^{2}k(0)Z+\frac{i\alpha\lambda}{\mu}Z_{\mu}-i\alpha \mu(\frac{Z_x}{\mu})_{\mu}
+\mu^2k_{xx}(0)Z,
\end{equation}
where we have assumed the statistical homogeneity $(Z_{x}=0)$.
Now we need the $\gamma$ which is given by $\gamma=\bar {h}_{t}$.
So invoking to the eq.(\ref{bur}) we get,
\begin{equation}{\label{gam}}
\gamma(t)=\bar{h}_{t}=\frac{\alpha}{2}\langle u^2\rangle.
\end{equation}
To evaluate $\langle u^2\rangle$ we set $\lambda=0$ in 
the eq.(\ref{gen-eq3}), and find,
\begin{equation}
Z_{t}=\mu^2k_{xx}(0)Z,
\end{equation}
which considering $Z(\mu,0)=1$ as initial condition, its solution is, 
\begin{equation}
Z(\mu,t)=\exp(\mu^{2}k_{xx}(0)t).
\end{equation}
On the other hand by definition we have
$\langle u^{2}\rangle=-(\frac{\partial^2 Z(\mu,t)}{\partial\mu^2})$.
So before creation of the singularities the second moment of 
height gradient behave as:
\begin{equation}
\langle u^2\rangle=-2k_{xx}(0)t,
\end{equation}
so consequently,  
\begin{equation}{\label{gam2}}
\gamma(t)=-\alpha k_{xx}(0)t.
\end{equation} 
Inserting eq.(\ref{gam2}) in the equation (\ref{gen-eq3}) results to,
\begin{eqnarray}{\label{gen-eq4}}
\frac{\partial}{\partial t} Z(\mu,\lambda,t)&=&-i\alpha\lambda 
{\frac {\partial ^{2}}{\partial \mu^{2}}}{Z}(\mu, \lambda, t)
+ i\alpha\frac{\lambda}{\mu}\frac {\partial }{\partial \mu}Z(\mu, \lambda, t)
\cr\nonumber\\
& & + (\mu^2 k_{xx}(0)-i\alpha k_{xx}(0)t\lambda - \lambda^{2} k(0))
Z(\mu, \lambda,t).
\end{eqnarray}
We solve the eq.(\ref{gen-eq4}) with the initial condition 
$Z(\mu,\lambda,0)$=1, from which by expanding the generating function in 
powers of 
$\lambda$ and $\mu$ we can obtain the moments 
$\langle (h-\bar {h})^{n}\rangle$ ,$\langle u^{n}\rangle$ 
and $\langle (h-\bar{h})^{n} u^{m}\rangle$. 
We change the variable $\mu$ to $y=\mu^{2}$, 
so the eq.(\ref{gen-eq4}) converts to the following equation,
\begin{eqnarray}
\frac{\partial}{\partial t} Z(y,\lambda,t)&=&-2i\alpha\lambda y
{\frac {\partial ^{2}}{\partial y^{2}}}{Z}(y, \lambda, t)
+ i\alpha\lambda\frac {\partial }{\partial y}Z(y, \lambda, t)\cr\nonumber\\
& & + (yk_{xx}(0)-i\alpha k_{xx}(0)t\lambda - \lambda^{2} k(0))
Z(y, \lambda,t).
\end{eqnarray}
Introducing the Fourier transform of $Z(\mu,\lambda,t)$ with respect to $y$
as $Q(q,\lambda,t)$, it is simple to get the following evolution equation 
satisfied by the Fourier transform, 

\begin{eqnarray}{\label{gen-eq5}}
\frac {\partial }{\partial t}Q(q, \lambda, t)&=&2\alpha \lambda
q^{2}\frac {\partial }{\partial q}Q(q,\lambda, t)+5\alpha\lambda q
Q(q, \lambda, {t})-ik_{xx}(0)\frac {\partial }{\partial
q}Q(q,\lambda, t)+
\nonumber\\
& & - i\alpha k_{xx}(0)t\lambda Q(q, \lambda, t) -\lambda^{2}k(0)
Q(q,\lambda, {t}),
\end{eqnarray}

with the initial condition  

\begin{equation}{\label{ini}}
Q(q, \lambda , 0)=\frac{1}{2\pi}\int e^{iyq}dy=\delta (q).
\end{equation}

The eq.(\ref{gen-eq5}) is a first order partial differential equation 
which can be solved by the method of characteristics. 
The general solution of eq. (\ref{gen-eq5}) is written as,

\begin{eqnarray}
&&Q(q,\lambda,t)=g\left(\lambda,\frac {1}{2}\frac {
2tk_{xx}(0)\alpha\lambda +\sqrt{-2ik_{xx}(0)\alpha\lambda}
\tanh^{-1}(q\sqrt{-\frac{2i\alpha\lambda}
{k_{xx}(0)}})}{k_{xx}(0)\alpha\lambda}\right)\times\nonumber\\
&&\times\exp\Bigg\{-1/2\int _{0}^{q}\Big(\frac {10\alpha\lambda s+ i 
\tanh^{-1}
(s\sqrt{\frac{-2i\alpha\lambda}{k_{xx}(0)}})\sqrt{-2ik_{xx}(0)\alpha\lambda}}
{2\alpha\lambda s^{2} - ik_{xx}(0)}
 \nonumber\\
&& \frac{+2i\alpha k_{xx}(0) t \lambda - 2\lambda^{2} k(0) -i
\tanh^{-1}(q\sqrt{\frac{-2i\alpha\lambda}{k_{xx}(0)}})
\sqrt{-2ik_{xx}(0)\alpha\lambda}} {2\alpha\lambda
s^{2}-ik_{xx}(0)}\Big)d s \Bigg \},
\end{eqnarray}

where $g$ is an arbitrary function of its arguments. 
Imposing the initial condition, given in eq.(\ref{ini}), 
and introducing 
$\omega$ as, 

\begin{eqnarray}
\omega =  {\frac {1}{2}} { \frac {
{\tanh^{-1}}(q\sqrt{\frac{-2i\alpha \lambda
}{k_{xx}(0}}))\sqrt{-2i k_{xx}(0)\alpha
\lambda}}{{{k}_{{xx}}}(0)\alpha \lambda }},
\end{eqnarray}

we obtain, 

\begin{eqnarray}
& &{g}(\lambda , \omega )=\delta ( - { \frac {1}{2}} \sqrt{2}\sqrt{ { \frac
{i{{k}_{{xx}}}(0)}{\alpha \lambda }} }{\tanh}(\sqrt{2i{
{k}_{{xx}}}(0)\alpha \lambda} \omega )) \exp [\! 1/2\int
_{0}^{-{\frac {1}{2}} \sqrt{2}\sqrt{
{\frac{i{{k}_{{xx}}}(0)}{\alpha \lambda }}
}{\tanh}(\sqrt{2i{{k}_{{xx}}}(0)\alpha \lambda} \omega )}\nonumber
\\
& &(\frac{10\alpha \lambda s + i\tanh^{-1}[\!
s\sqrt{\frac{-2i\alpha\lambda}{k_{xx}(0)}} \! ]
\!\sqrt{-2ik_{xx}(0)\alpha\lambda}-2i\omega
k_{xx}(0)\alpha\lambda - 2\lambda ^{2} k(0)}
{2\alpha \lambda s^{2} - ik_{xx}(0)})ds \!  ],
\end{eqnarray}

from which $Q(q,\lambda,t)$ is obtained as,

\begin{eqnarray}{\label{gen-s1}}
& &Q(q, \lambda , {t})= {g}(\lambda , {t} + \omega ) \exp [\!
-1/2\int _{0}^{q}\frac {10\alpha \lambda s + i\tanh^{-1}[\!
s\sqrt{\frac{-2i\alpha\lambda}{k_{xx}(0)}} \! ]
\!\sqrt{-2ik_{xx}(0)\alpha\lambda}}{2\alpha \lambda s^{2} - ik_{xx}(0)} 
\nonumber\\
& &\frac{-2i\alpha\lambda k_{xx}(0)t-i\tanh^{-1}[\!
q\sqrt{\frac{-2i\alpha\lambda}{k_{xx}(0)}} \! ]
\!\sqrt{-2ik_{xx}(0)\alpha\lambda} - 2\lambda ^{2} k(0)}
{2\alpha \lambda s^{2} - ik_{xx}(0)}ds \! ] . 
\end{eqnarray}

Inverse Fourier transforming of the solution in eq.(\ref{gen-s1}) is 
straightforward, so after switching 
to variable $\mu$ we get the following solution for $Z(\mu,\lambda,t)$:
\bea{\label{gen-s2}}
Z(\mu, \lambda , {t})&=& \Big( 1 -
{\tanh}^{2}(\sqrt{2i{{k}_{{xx}}}(0)\alpha \lambda}{t})\Big)
\exp\Big\{- \frac{5}{8}\ln(1-\tanh^{4}(\sqrt{2i{{k}_{{xx}}}(0)\alpha 
\lambda}{t})) \cr \nonumber \\
&+&\frac{5}{4}\tanh^{-1}(\tanh^{2}(\sqrt{2ik_{xx}(0)\alpha\lambda}t))
-\lambda^{2}k(0)t \cr \nonumber \\
&-&\frac{1}{16}\ln^{2}(\frac{1-\tanh(\sqrt{2ik_{xx}(0)\alpha\lambda}t)}
{1+\tanh(\sqrt{2ik_{xx}(0)\alpha\lambda}t)})
-\frac{1}{2}i\mu^{2}\sqrt{\frac{2ik_{xx}(0)}{\alpha\lambda}}{\tanh}
(\sqrt{2i{{k}_{{xx}}}(0)\alpha \lambda}{t})\Big\}.
\eea
 
Since we are interested in moments $\langle (h-\bar{h})^{n}\rangle$, so 
setting $\mu=0$ in (\ref{gen-s2})  
and expanding the generating function in power of $\lambda$ we can obtain 
them all. 
For example expanding up to $O(\lambda^6)$ it is easy to see that
the first sixth order of moments are behaving as following,
\begin{eqnarray}
&&\langle (h-\bar{h})^2\rangle=-\frac {1}{3}t(k_{xx}(0)^{2}\alpha ^{2}t^{3}
-6{k}(0)), \\
&&\langle (h-\bar{h})^3\rangle=-\frac {24}{45}k_{xx}(0)^{3}\alpha ^{3}t 
^{6},\\ 
&&\langle (h-\bar{h})^4\rangle=-\frac{101}{105}k_{xx}(0)^{4}\alpha
^{4}{t}^{8}-\frac{1}{6}t^{5}k_{xx}(0)^{2} \alpha ^{2}{k}(0)
+\frac{1}{2}t^{2}
k(0)^{2},\\
&&\langle (h-\bar{h})^5\rangle=-(\frac{2288}{945}k_{xx}(0)^{5}\alpha
^{5}t^{10} + \frac{4}{45} k_{xx}(0)^{3}\alpha^{3}t^{7}k(0)), \\
&&\langle (h-\bar{h})^6\rangle=-\frac
{1}{10395}{t}^{3}(85783k_{xx}(0)^{6}\alpha
^{6}t^{9} + 299970 k_{xx}(0)^{4}\alpha ^{4}t^{6}k(0)\hspace{3cm}\nonumber \\
&& + 623700 k_{xx}(0)^{2}\alpha ^{2}t^{3}k(0)^{2} - 1247400
k(0)^{3}).
\end{eqnarray}

The important content of the exact form derived above is that through 
them the time scale of sharp valley formation can be found.  
Actually there is no guarantee that the following generating function can 
be derived from a physical probability density function. So one should first check 
the realisability condition, i.e. $P(h-{\bar h},t) > 0$. 
In fact the above moment relations indicate 
that different even order moments become {\it negative} in some distinct 
characteristic time scales.
Closer looking in the even moment relations reveals that
the higher the moments are, the smaller their characteristic time scales 
become. So asymptotically the rate of decrease tends to $\frac{1}{4}t^{*}$ 
for very large even moments, where 
$t^{*}=(\frac{k(0)}{\alpha^2 k_{xx}^{2}(0)})^{1/3}$ ( see figure 1). 
Therefore we conclude that after this time the far tails 
of the probability distribution function start to become negative, 
which is reminiscent of sharp valley creation. It means that after the 
characteristic time scale $t^{*}$ one should also consider the 
contribution of the relaxation term in the limit of vanishing diffusion
in order to find a realisable probability density function of height field.
In other words disregarding the diffusion 
term in the PDF equation is valid only up to the time scales in 
which the singularities are developed. 
Taking into account that $\alpha > 0$, the odd order 
moments are positive in time scales before formation of sharp valleys.
It means that the probability density $P(h-{\bar h},t)$ in this time 
regime is negatively skewed.
In figures [3-5] we have demonstrated the role of $\sigma$ on the time
scale of creation of singularities. Substituting the $k_{xx}(0)$ and 
$k(0)$ in the expression of $t^{*}$ it gives us $t^* = (\pi)^{1/6} {D_0}^{-1}
\alpha^{-2/3} \sigma^{5/3} $. Hence the smaller the $\sigma$ is
the shorter the time scale of shock creation would be(see figures 3,4 and 5).

\section{ The equation of joint PDF of height difference and height 
gradient in the stationary state}

Assuming a stationary state,
we are interested in investigating the stationary solutions of 
eq.(\ref{pdf-eq2})
in the limit $\nu \rightarrow 0$. Of course in the stationary state the sharp 
valleys 
are fully developed and one should also take care of the diffusion term 
in the PDF equation.
The complicated term involved with the singularities
, can be cured by using the method proposed in [73].
Let us define, 
\bea{\label{anom1}}
G(u,{\tilde h},t)&& =\lim_{\nu \rightarrow 0} \nu \langle u_{xx}|u,{\tilde h}\rangle
P(u,{\tilde h},t) \cr \nonumber \\
&& =\lim_{\nu \rightarrow 0} \nu \langle u_{xx}(x,t) 
\delta(\tilde h - \tilde h(x,t))\delta(u-u(x,t))\rangle ,
\eea
,where in the last step in eq.(\ref{anom1}) we have used the definition of 
joint PDF $ P(u,{\tilde h},t)$. 
Assuming spatial ergodicity, the average of the dissipative term can be  
expressed as,

\bea{\label{anom2}}
&&\nu \langle u_{xx}|u,{\tilde h}\rangle P(u,{\tilde h},t)=
\nu\langle u_{xx}(x,t)\delta(u-u(x,t))
\delta({\tilde h}-{\tilde h}(x,t)\rangle \nonumber \\
&&=\nu\lim_{L\rightarrow\infty}\frac{N}{L}\frac{1}{N}\int_{-L/2}^{L/2}
dx u_{xx}(x,t)\delta(u-u(x,t))\delta({\tilde h}-{\tilde h}(x,t)).
\eea

It is well-known that the $u$-field which satisfies the Burgers equation, 
gives rise to discontinuous or shock solutions in the limit $\nu \rightarrow 
0$. Consequently for finite $\sigma$ the shock  
solutions are manifested in height field as a set of sharp valleys 
at the positions where the shocks are located, 
where they are continuously connected by some hill configurations 
Fig(5). 
It is noted that $u_{xx}$ is zero at the positions where no sharp valley exists. 
Therefore in the limit $\nu\rightarrow 0$ only small intervals around 
the sharp valleys will contribute to the integral in the eq.(\ref{anom2}). Within 
these intervals, boundary layer analysis 
can be used for obtaining accurate approximation of $u(x,t), {\tilde h}(x,t)$.  
Generally boundary layer analysis deals with the problems in which 
perturbations are operative over very narrow regions across which the 
dependent variables undergo very rapid changes. 
These narrow regions (sharp valley layers) frequently  
adjoin the boundaries of the domain of interest, owing the fact that a
small parameter ($\nu$ in the present problem) multiplies the highest 
derivative.
A powerful method for treating boundary layer problems is the method of
matched asymptotic expansions.
The basic idea underlying this method is that an approximate solution to a
given problem is sought not as a single expansion in terms of a single scale,
but as two or more separate expansions in terms of two or more scales
each of which is valid in part of the domain. The scales are chosen, so that
the expansion as a whole, covers the whole domain of interest and the 
domains  of validity of neighbouring expansions overlap.
In order to handle the rapid variations in the sharp valley layers, we define a  
suitable magnified or stretched scale and expand the functions in terms of it
in the sharp valley regions.
For this purpose, we split $u$ and $h$ into a sum of 
inner solution near
the sharp valleys and an outer solution away from the sharp valleys, and use systematic
matched asymptotics to construct uniform approximation of $u$ and $\tilde h$.
For the outer solution, we look for an approximation in the form of a series
in $\nu$,
\bea
u&&=u^{out}=u_{0}+\nu u_{1}+O(\nu^{2}), \cr \nonumber \\
{\tilde h}&&={\tilde h}^{out}={\tilde h}_{0}+\nu {\tilde h}_{1}+O(\nu^{2}),
\eea
where $u_0$ and $ {\tilde h}_{0}$ satisfy the Burgers and KPZ equation 
without dissipation term,
\bea
&&u_{0t} +\alpha u_0 u_{0x} =- f_x, \cr \nonumber \\
&& {\tilde h}_{0t} - \frac{\alpha}{2} (\partial_x {\tilde h}_{0})^2 = f.
\eea

In order to deal with the inner solution around the sharp valley , let $y = y(t)$
be the position of a sharp valley, define the 
stretched variable $z=\frac{x-y}{\nu}$ and let,

\bea
u^{in}(x,t)&=&v(\frac{x-y}{\nu}+\delta,t),\cr \nonumber \\
{\tilde h}^{in}(x,t)&=&{\tilde h}(\frac{x-y}{\nu}+\delta,t).
\eea

The parameter $\delta$ is a perturbation of the sharp valley position and $v$ and 
$\tilde h$ satisfy the following equations,
\bea
&&\nu v_t-\alpha({\bar u}-\nu\eta)v_z+\alpha vv_z=v_{zz}-f_{z}(z,t),\cr 
\nonumber \\
&&\nu^2 {\tilde h}_t-\nu {\bar u} {\tilde h}_z+\eta\nu^2 {\tilde h}_z
-\frac{\alpha}{2}({\tilde h}_z)^2=\nu {\tilde h}_{zz}+\nu^2 f(z,t).
\eea

${\bar u}=\frac{1}{\alpha}{\frac{dy}{dt}}=\frac{u_{+}+u_{-}}{2},
\eta=\frac{1}{\alpha}\frac{d\delta}{dt}$ and $u_{+}, u_{-}$ are the height 
gradients in right hand and left hand sides of the sharp valley in the 
position $y$, ( see figure-2).
We look for a solution in the form,

\bea
v&=&v_{0}+\nu v_{1}+O(\nu^{2}), \cr \nonumber \\
{\tilde h}&=&{\tilde h}_{0}+\nu {\tilde h}_{1}+O(\nu^{2}).
\eea

To leading order we get for $v_{0}$ and ${\tilde h}_{0}$,
                                                                             
\bea{\label{boun}}
&&{\tilde h}_{0z}=0, \cr \nonumber \\
&&\alpha(v_{0}-\bar u)v_{0z}={v_0}_{zz},
\eea
where we have assumed that variance of $f(z,t)$ is a smooth function 
so that we can neglect its variation in the sharp valley region ($f_z=0$).
In other words we suppose that $\sigma >> O (\nu) 
(i.e. \sigma >>$ the typical layer width).
One can easily integrate the eq.(\ref{boun}) and find that,
\bea
&&{\tilde h_0} ={\rm const}, \cr \nonumber
&&v_{0}={\bar u}-\frac{s}{2}\tanh(\frac{\alpha sz}{4}), 
\eea
in which $s=s(t)=u_{+}-u_{-}$ is the shock strength.
The boundary condition for this equation arises from the matching condition,

\be
\lim_{z\rightarrow {\pm\infty}} v_{0}^{in}=\lim_{x \rightarrow y} u_{0}^{out}=
{\bar u}{\pm\frac{s}{2}}.
\ee
Basically since $ {\tilde h}_{0}^{in}(z)= C -\nu \int^{z} v_{0}^{in}(z') 
dz' $, where $C$ is the integration constant.
So the O(1) solutions of $v_{0}^{in}$ give rise to O($\nu$) solutions in 
${\tilde h}_{0}^{in}$ field and only the integration constant is the O(1) 
part of the solution of ${\tilde h}_{0}^{in}$. In fact the constant in 
nothing but the height value at the 
sharp valley position. Of course due to the height continuity at sharp valley 
position 
there is no boundary layer for KPZ equation meaning that the rapid 
changing term in the sharp valley layer occurs in $h_{xxx}$ while the highest 
derivative in KPZ equation involves only $h_{xx}$. 
The above analysis show that, to $O(\nu)$, eq.(\ref{anom2}) can be 
estimated as,

\bea
&& \lim_{\nu \rightarrow 0 } \nu (\langle u_{xx}|u,{\tilde h}\rangle P 
(u,{\tilde h},t)), \cr \nonumber \\ &&=
\lim_{\nu \rightarrow 0 } \nu\lim_{L\rightarrow\infty}\frac{N}{L}\frac{1}{N}\sum_j\int_{{\Omega}_j}
dx u_{xx}^{in}\delta(u-u^{in}(x,t))\delta({\tilde h}-{\tilde 
h}^{in}(y_{j},t)), \cr \nonumber \\
&& =
\lim_{L\rightarrow\infty}\frac{N}{L}\frac{1}{N}\sum_{j}\int_{-\infty}^{\infty}
dz u_{zz}^{in}\delta(u-u^{in}(z,t))\delta({\tilde h}-{\tilde 
h}^{in}(y_{j},t)), \cr \nonumber \\
&& = \lim_{L\rightarrow\infty}\frac{N}{L}\frac{1}{N}\sum_{j}\int_{-\infty}^{\infty}
dz {v_{0}}_{zz}^{in}\delta(u-v_0)\delta({\tilde h}-{\tilde h}(y_{j},t)),
\eea

where $\Omega_{j}$ is a layer located at $y_{j}$ with width $>>O(\nu)$.
Using the eq.(\ref{boun}) and,
\be
dz {v_0}_{zz}=dv_0\frac{{v_0}_{zz}}{{v_0}_z}=\alpha dv_0(v_0-\bar u),
\ee
the $z-$integral can be evaluated exactly leading to the following result,
\be{\label{anom3}}
\nu \langle u_{xx}|u,{\tilde h}\rangle P (u,{\tilde h},t)=
\alpha\int d{\bar u}\int_{-\infty}^{0} ds \varrho({\bar u},s,{\tilde h},x,t)
\int_{{\bar u}+
\frac{s}{2}}^{{\bar u}-\frac{s}{2}}dv_{0}(v_{0}-{\bar u})\delta(u-v_{0}).
\ee
$\varrho({\bar u},s,{\tilde h},x,t)$ is defined such that 
$\varrho({\bar u},s,{\tilde h},t)d{\bar u}dsd{\tilde h}dx$ gives the average
number of vallies in $[x,x+dx)$ with ${\bar u}(y,t)\in[{\bar u},{\bar u}+d{\bar u})$,
$s(y,t)\in[s,s+ds)$ and ${\tilde h}(y,t)\in[{\tilde h},{\tilde h}+d{\tilde h})$,
where $y\in[x,x+dx)$ is the sharp valley location. 
Eq.(\ref{anom3}) indicates that the relaxation term in the strong coupling 
limit can be written in terms of
some quantities which are defined in singularities (vallies). Indeed we 
characterise a sharp valley with four quantities, its location $y_j$, its 
gradients at 
$y_{j_{0^+}}$ (i.e. $u_+$),  $y_{j_{0^{-}}}$ (i.e. $u_-$) and its height from
the $\bar h$ i.e. $\tilde h_j$. Instead of $u_+$ and $u_-$ we have used the 
quantities $ {\bar u}=\frac{u_{+}+u_{-}}{2}$ and $s=s(t)=u_{+}-u_{-}$.
Later we will determine the time evolution equations which govern over 
these four quantities.

Proceeding further we note that $\varrho({\bar u},s,{\tilde h},x,t)$ can be 
defined as,

\be
\varrho({\bar u},s,{\tilde h},x,t)=\langle\sum_j \delta({\bar u}-{\bar u}(y_j,t) 
\delta(s-s(y_j,t))\delta({\tilde h}-{\tilde h}(y_j,t))\delta(x-y_j)\rangle.
\ee
Due to statistical homogeneity the sharp valley's characteristics 
are independent of their location, so

\be 
\varrho({\bar u},s,{\tilde h},x,t)=\rho S({\bar u},s,{\tilde h},t)),
\ee
in which $\rho=\rho(t)$ is the number density of shocks and
$S({\bar u},s,{\tilde h},t)$ is the PDF of $({\bar u}(y_{0},t),s(y_{0},t),
{\tilde h}(y_{0},t))$ conditional on $y_{0}$ being a shock location. 
Hence 

\be
\lim_{\nu\rightarrow 0}\nu \langle u_{xx}|u,{\tilde h}\rangle P (u,{\tilde h},t)=
-\alpha\rho\int_{-\infty}^{0} ds\int_{u+\frac{s}{2}}
^{u-\frac{s}{2}}d{\bar u}(u-{\bar u})S({\bar u,s,{\tilde h},t}).
\ee

Therefore the relaxation (dissipative) contribution in eq.(\ref{pdf-eq2})
is written as,

\be{\label{anom4}}
G(u,{\tilde h},t)=-(\alpha\rho\int_{-\infty}^{0} ds\int_{u+\frac{s}{2}}
^{u-\frac{s}{2}}d{\bar u}(u-{\bar u})S({\bar u,s,{\tilde h},t}))_{uu}.
\ee
So eq.(\ref{pdf-eq2}) is rewritten in the following form,
\be{\label{pdf-eq3}}
-P_{ut}=-\gamma P_{{\tilde h}u}-\frac{\alpha}{2}(u^2P)_{{\tilde h}u}
-\alpha(uP)_{\tilde h}-k(0)P_{{\tilde h}{\tilde h}u}+k_{xx}(0)P_{uuu}+
G(u,{\tilde h},t).
\ee
It is interesting that the the $G$-term comes from the relaxation term
in the KPZ equation, but its explicit expression in terms of sharp valley's
characteristics is proportional to $\alpha$ which is the coefficient of
the nonlinear term in the KPZ equation. This indicates that without
the nonlinear term in the KPZ equation there is no finite contribution 
for the diffusion term in the PDF equation when $\nu \rightarrow 0$. 
Although this equation is exact for finite $\sigma$ however we can not  
solve it since the last term is not expressed in 
terms of $P(u,{\tilde h},t)$. 
Despite the existence of unclosed $G$-term still we can derive interesting 
information the moments using the above equation.
we will
study comprehensively the moments of height difference and height gradient, 
i.e. $\langle (h -\bar h)^n (\partial_x h)^m \rangle $ in the next section. 

But before it's worth of remarking that integration over $\tilde h$ gives an 
equation for probability density function (PDF) of $u$ recovering the 
results in [73], 
\be{\label{req1}}
R_{t} =- k_{xx}(0) R_{uu} + \{ \rho \alpha \int_{- \infty} ^0 
\int_{u+\frac{s}{2}} ^{u-\frac{s}{2}} d \bar u  (u-\bar u) S (\bar u,s,t)\}_u 
\ee
where $R (u,t) = \int P(u, \tilde h,t) d \tilde h$ and 
$S ( \bar u,s,t) = \int S ( \bar u, s, \tilde h,t )  d \tilde h$
is the PDF of $(\bar u(y_0,t), s(y_0,t))$, conditional on
the property that $y_0$ is the shock (cusp) position. Because of the 
statistical homogeneity, $y_0$ is a dummy variable.
We finish the section with determining the relation between 
the density of vallies and the noise characteristics $k (0)$ and 
$  k_{xx}(0) $, that is,
\be{\label{kxx}}
k_{xx}(0)=\frac{\alpha \rho}{12}\langle s^3 \rangle
\ee
indicating that the forcing variance $ k_{xx}(0)$ is related
to the products of density of vallies $\rho$, $ \langle s^3 \rangle$
and $\alpha$. 
This relation has been found in [73] 
and its details are given in the appendix. 

\section{ The moments of height fluctuation in the stationary state}

Our goal is investigating the scaling behaviour of moments 
of height difference and height gradient in the stationary state. 
After sharp valley formation the lateral correlations produced by 
nonlinear term will grow with time. Dynamic scaling 
exponent $z$ characterises the self-similar growth of the lateral growth. 
However in the stationary state the height field width saturates in the 
sense that lateral correlations are in average grown up to the system 
size. As it was explained, after the saturation the width scales as 
$w_0(L,t\gg L^{z}) \sim L^{\chi}$. Having in our disposal the exact result 
$\chi=\frac{1}{2}$ in one dimension [89], it would be natural to define 
$P(h',u,t)$   
as PDF of $h'$, $u$, and $t$, where $h'= \frac {h-\bar h}{w_0}$
and $w_0= L ^{1/2}$. Obviously
$P(h',u,t)$ is related to $P({\tilde h},u,t)$ as $P(h',u,t)=w_0 P({\tilde 
h},u,t)$.
From eq.(\ref{pdf-eq3}) it follows that $P(h',u,t)$ in the stationary state 
satisfies the following equation,
\bea{\label{pdf-st}}
&&-\gamma L^{-1/2} P_{h'u}+
-\frac{\alpha}{2} L^{-1/2}(u^2 P)_{h'u}
-\alpha L^{-1/2}(uP)_{h'} \cr \nonumber \\
&& -k(0) L^{-1} P_{h'h'u} + k_{xx}(0)P_{uuu} + G(u,h',t) = 0.
\eea
From eq.(\ref{pdf-st}) it follows that the moments of $\langle h'^n 
u^m\rangle$ satisfies the following equation in the stationary state,
\bea{\label{mom-st1}}
g_{n,m} &+& k_{xx}(0)m(m-1)(m-2)\langle h'^{n}u^{m-3}\rangle  
+ mn(n-1)k(0)L^{-1}\langle h'^{n-2}u^{m-1}\rangle \cr \nonumber \\
&-&mn\gamma L^{-1/2}
\langle h'^{n-1}u^{m-1}\rangle  
-n(m-2)\frac{\alpha}{2}L^{-1/2}\langle h'^{n-1}u^{m+1}\rangle = 0,
\eea
where
\be
g_{n,m}=\int dh'\int du h'^{n}u^{m} G(u,h',t).
\ee
Using the eq.(\ref{anom4}), 
we can determine the explicit expression of $g_{n,m}$ in terms of 
the characteristics of vallies. 
Thus we find,
\be
g_{n,m} =-\alpha \rho \int d h' \int du h'^n u^m 
(\int_{-\infty}^{0} ds\int_{u+\frac{s}{2}}
^{u-\frac{s}{2}}d{\bar u}(u-{\bar u})S({\bar u,s,h',t}))_{uu}.
\ee
After integrating by parts it converts to,
\bea
g_{n,m} &&=- \alpha \rho \int d h' \int du h'^n  m(m-1) u^{m-2} 
(\int_{-\infty}^{0} ds\int_{u+\frac{s}{2}}
^{u-\frac{s}{2}}d{\bar u}(u-{\bar u})S({\bar u,s,h',t})),\cr \nonumber \\
&&=- \alpha \rho  m (m-1) \int_{-\infty} ^0 ds \int d h' h'^n \int du u^{m-2} 
(\int_{u+\frac{s}{2}}
^{u-\frac{s}{2}}d{\bar u}(u-{\bar u})S({\bar u,s,h',t})).
\eea
It can be integrated over $u$ which leads to the following expression for 
$g_{n,m}$ (see appendix B),

\bea{\label{gnm}}
g_{n,m}&&=\frac{\alpha\rho}{2^m}(\langle h_v '^{n}(2{\bar u}+s)^{m-1}
[(m-1)s-2{\bar u})]\rangle\cr \nonumber\\
&&+\langle h_v'^{n}(2{\bar u}-s)^{m-1}[(m-1)s+2{\bar u})]\rangle),
\eea

where $h_v ' = \frac{h_v - \bar h}{w}$ and $h_v$ is the height of given sharp valley .
This means that the relaxation term in the strong coupling
limit can be written in terms of $only$ characteristics
of the vallies i.e. $\bar u$, $s$ and $h_v$.

At statistical steady state ($t\rightarrow\infty,\langle \ldots\rangle_t=0$),
and assuming the scale independence of $g_{n,m}$'s, one can derive from 
eq.(\ref{mom-st1}), to leading order in the limit of $L\rightarrow 
\infty$ the following, 
\be{\label{mom-st2}} 
\langle (\frac {h - \bar h }{w_0})^n u^{m-3}\rangle =
-\frac{g_{n,m}}{k_{xx}(0)m(m-1)(m-2)},
\ee                                                                        
for $m \geq 3$.
Putting $n=0$ then for instance results in the height 
gradient moments behaving as [73],

\bea{\label{u-mom}}
\langle u^{m}\rangle&=&\frac{\alpha\rho}{2^{m+3}k_{xx}(0)(m+1)(m+2)(m+3)}\cr \nonumber \\
& &\{\langle (2{\bar u}-s)^{m+2}[(m+2)s+2{\bar u})]\rangle-
\langle (2{\bar u}+s)^{m+2}[2{\bar u}-(m+2)s]\rangle\}.
\eea
and for $m=3$ we find,
\be {\label{h-mom1}}
\langle ( h - \bar h)^n \rangle = \frac {g_{n,3}}{6 k_{xx} (0)} 
\hskip .2cm    L^{n/2}
\ee

where using the eq.(70) we have  $ g_{n,3} = \frac{\alpha\rho}
{2}\langle h^{'n}s^{3}\rangle$.
Assuming that $g_{nm}$'s are scale independent, at least in leading order, 
then one may argue that eq.(\ref{h-mom1}) builds up a relation between the 
$n$-th moments of height difference in term of the second moment
$w_0$ in a non-intermittent way. That is the $n$-th order moment is scaled 
linearly with order $n$. 
Rationalising the assumption of $g_{n,m}$'s scale independence, one would 
look at the statistics of
sharp valley environment and the different processes 
involved in the sharp valley creation 
and annihilation which contribute dynamically. We will postpone 
the argumentation to next section and instead for the rest of this 
section we find out the consequences of scale independency of $g_{n,m}$s. 
In the end of this section we will verify the requirements of positivity 
and normalisability of the joint $P({\tilde h},u,t)$ in the stationary state.

Eq.(\ref{mom-st1}) also suggests that the amplitudes of height difference 
and height gradient
moments depend strongly on the singular structures in the theory, encoded
in the functions $g_{n,m}$ (i.e. eq.(\ref{gnm})).

At this stage we find the finite size effects on the 
moments of $S_{n,m}=\langle h'^{n}u^{m}\rangle$.
Defining $\epsilon = 1/L^{1/2}$ as a perturbative parameter 
we find the structure functions
$S_{n,m}=\langle h'^{n}u^{m}\rangle$ 
perturbatively in terms of perturbative parameter 
$\epsilon = \frac {1} {L^{1/2}}$ as ,
\be{\label{snm-def}}
S_{n,m}=\langle h'^{n}u^m\rangle=S_{n,m}^{(0)}+\epsilon S_{n,m}^{(1)}
+{\epsilon}^2 S_{n,m}^{(2)}+\cdot\cdot\cdot,
\ee
Invoking to eq.(\ref{snm-def}) and assuming the scale independency of 
$g_{nm}$ we get,

\be
S_{n,m}^{(0)}=\frac{1}{k_{xx}(0)(m+1)(m+2)(m+3)}g_{n,m+3},
\ee

\bea
S_{n,m}^{(1)}&&=\frac{1}{k_{xx}(0)(m+1)(m+2)(m+3)}\{ \frac{\gamma n(m+3)}
{k_{xx}(0)(m+3)(m+4)(m+5)}g_{n-1,m+5} \cr \nonumber 
&&+\frac{\alpha n(m+1)}{2k_{xx}(0)(m+5)(m+6)(m+7)}g_{n-1,m+7} \},
\eea
and, 
\bea
S_{n,m}^{(2)}&&=\frac{1}{k_{xx}(0)(m+1)(m+2)(m+3)}
[{\gamma n(m+3)}S_{n-1,m+4}^{(1)} \cr \nonumber \\
&&\frac{\alpha}{2}n(m+1)S_{n-1,m+4}^{(1)}-k(0)n(n-1)(m+3)S_{n-2,m+2}^{(0)}],
\eea
and etc. For example
the moments $ \langle ( h - \bar h )^n \rangle$ behave as,
\bea{\label{h-mom2}}
&& \langle ( h - \bar h )^n \rangle =  L^{n/2}  \hskip .2 cm \{ \frac{1}{3! k_{xx}(0)} g_{n,3}
 \cr \nonumber \\ 
 &&+ \frac {1}{L^{1/2}} (   -\frac{\gamma n}{5! k_{xx}(0) ^2 } g_{n-1,5} -    
 \frac{2 \alpha n }{7! k_{xx}(0) ^2 } g_{n-1,7}) \cr \nonumber \\
 &&+ \frac {1}{L} (  \frac{k(0) n (n-1) }{5! k_{xx}(0) ^2 } g_{n-2,5} 
 + \frac{\gamma^2 n (n-1)}{7! k_{xx}(0) ^3} g_{n-2,7} \cr \nonumber \\
 &&-\frac{11 \alpha \gamma n (n-1) }{9! k_{xx}(0)^2 } g_{n-2,9} 
  -\frac{40 \alpha^2 n (n-1) }{11! k_{xx}(0)^2} g_{n-2,11})+ O( L^{-3/2}) \} 
\eea
Noting to the fact that the $g_{n-1,5}$ and $g_{n-1,7}$ are not zero,
therefore we conclude that the next to leading order correction 
for structure functions is $O(1/{L^{1/2}})$.
Also eq. (\ref{h-mom2}) shows that the amplitude of the correction terms to 
moments 
$  \langle ( h - \bar h )^n \rangle $ are related to the statistics 
of quantities which are defined on the singularities i.e. $g_{n,m}$. 
Also it shows all of the moments $ \langle ( h - \bar h )^n \rangle$ 
(for even and odd n) exist and consequently the PDF of the $h- \bar h $ is 
not symmetric. However using the 
properties of the Burgers equation, it can be shown that only even 
moments of $u$ are nonzero and all the odd moments vanish, hence the 
PDF of $u$ is symmetric.\\
Eq.(\ref{u-mom}) enables us to determine the rate of growth
of surface at the stationary state, i.e. $\bar h _t$. 
Using the KPZ equation one it is trivial to see that, 
\be
\lim_{t\rightarrow\infty}\gamma(t) ={\bar h}_t=
\frac{\alpha}{2}\langle u^2\rangle + \nu \rho\langle s \rangle,
\ee
where we have used the fact that $ \langle h_{xx} \rangle = - \langle u_x 
\rangle = - \rho \langle s \rangle $ [73].
In the limit $\nu \rightarrow 0$ the second term vanishes and,
\be
\lim_{t\rightarrow\infty}\gamma(t)={\bar h}_t=
\frac{\alpha}{2}\langle u^2 \rangle.
\ee

Back to the eq.(\ref{u-mom}), the  ${\bar h}_t$ is written in terms of 
properties of singularities as,
\be
{\bar h}_t = \frac {\alpha^2 \rho } {2^4 5! k_{xx}(0)} 
\{ \langle 80 \bar u ^2 s^3 + 4 s^5\rangle \}
\ee
So in the stationary state, 
moments $ \langle \bar u ^2 s^3 \rangle$ and $ \langle s^5 \rangle$ 
would determine the growth rate.
In other words for a given time in the steady state if one average the moments
$ \langle \bar u ^2 s^3 \rangle$ and $ \langle s^5 \rangle$, which 
is defined only on the vallies, can predict the rate of growth of surface.
This provides a simple way to determine the $ \bar h _t$ in the stationary
state.
Now we prove that $P(h',u,t \rightarrow \infty, L \rightarrow \infty )
 = P(h',u)$ is a positive
and normalisable PDF. Regarding the positivity of the PDF we note 
that eq.(\ref{req2})
indicates that $P(h',u)$ satisfies the following equation in the limit
of $L\rightarrow \infty$,
\be
k_{xx}(0)P_{uuu} = 
(\alpha\rho\int_{-\infty}^{0} ds\int_{u+\frac{s}{2}}
^{u-\frac{s}{2}}d{\bar u}(u-{\bar u})S({\bar u,s,{\tilde h},t}))_{uu}.
\ee
Invoking to the method introduced in section 3, one may obtain,
\be
P(h',u) = 
-\frac{ \alpha  \rho}{2k_{xx}(0)}\int_{-\infty}^{0} ds\int_{u+\frac{s}{2}}^
{u-\frac{s}{2}} 
d\bar {u} [\frac{s^2}{4}-(u-\bar {u})^2]S_{\infty}(\bar {u},s, h'),  
\ee
where $  S_{\infty}(\bar u,s,h') $ is the PDF of vallies with $\bar u$, $s$
and $h'$. Therefore positivity $ S_{\infty}(\bar u,s,h') \geq 0$ 
implies that $ P(h',u) \geq 0$.
To check the normalisability of the $P(h',u)$, consider
the eq.(\ref{mom-st2}) with $n=0,m=3$, leading to,

\be
\int d h' du P( h',u,t) =  \frac{1}{3! k_{xx}(0)} \hskip .2cm g_{0,3},
\ee

while explicit form of $g_{0,3}$ can be found from the eq.(\ref{gnm}).
The result
$g_{0,3} = \frac{\alpha\rho}{2} \langle s^3 \rangle $ cooked up with the 
eq.(\ref{kxx}) gives, 
\be
\int d h' du P( h',u,t\rightarrow \infty, L \rightarrow \infty) = 1
\ee
, completing the proof of normalisability of the stationary state PDF  
$P(h',u)$.

\section{statistics for the environments of the singularities}  

In this section we derive the PDF of quantities which characterise the
sharp valleys. As is depicted in Fig.(2) and formerly is described, the 
evolution of the surface after the formation of singularities 
is determined by the dynamics of the sharp vallies and their statistical 
properties. In a more quantitative sense one should attempt in  
characterising the time 
evolution of $h_{v_{j}}$, $\bar u$ 
and $s$ consequently. 
From there we show that to leading order of the expansion in terms of system 
size the $g_{n,m}$`s do not depend on the scale $L$. 
Doing so we reach to such a level of describing 
the dynamics of the surface growth by which one may also trace the 
dynamics of the singularity enviornments. That makes a logical way to
construct the pathway towards examining the statistical properties of 
singularity functions $g_{n,m}$. The importance of such an analysis 
became clear in the last section where the determination of the finite size 
corrections to scaling in $S_{n,m}=\langle h'^n u^m \rangle$ showed to be 
depended on the existence or lack of scale dependencies in the 
singularity functions $g_{n,m}$. \\ 
Let us turn to study the statistics for 
the environment of 
the singularities in KPZ equation. Defining $\xi(x,t)=-h_{xx}(x,t)$ and 
let $W(h_{v},\bar u,s,\xi_{+},\xi_{-},x,t)$ be the 
PDF of 

\begin{eqnarray*}
&&h_{v}(x,y_{j},t)=\frac{1}{2}(h(y_{j}+\frac{x}{2})+h(y_{j}-\frac{x}{2})),\\
&&{\bar u}(x,y_{j},t)=\frac{1}{2}(u(y_{j}+\frac{x}{2})+u(y_{j}-\frac{x}{2})),\\
&&s(x,y_{j},t)=u(y_{j}+\frac{x}{2})-u(y_{j}-\frac{x}{2}),\\
&&\xi_{\pm}(x,y_{j},t)=-h_{x_{\pm}x_{\pm}}(y_{j}\pm x_{\pm},t),
\end{eqnarray*}

conditional on $y_{j}$ being a singularity position.
In this section we will find the master equation governing the evolution of  
$W(h_{v},\bar u,s,\xi_{+},\xi_{-},x,t)$ in the limit of $\nu\rightarrow 0$.
Starting from the dynamical equation,

\begin{eqnarray}
& 
&h_{t}(z+x_{\pm})-\frac{\alpha}{2}h^{2}_{x_{\pm}}(z+x_{\pm})=f(z+x_{\pm}),\\
& &u_{t}(z+x_{\pm})+\alpha uu_{x_{\pm}}(z+x_{\pm})=-f_{x_{\pm}}(z+x_{\pm}),\\
& &\xi_{t}(z+x_{\pm})+\alpha u(z+x_{\pm})\xi_{x_{\pm}}(z+x_{\pm})+\alpha^{2}\xi
^{2}(z+x_{\pm})=-f_{x_{\pm}x_{\pm}}(z+x_{\pm}),
\end{eqnarray}

we define,
\bea
&&\theta(\lambda_{+},\lambda_{-},\mu_{+},\mu_{-},\eta_{+},\eta_{-},x_{+},
x_{-},z,t)=  \nonumber\\
&&\exp(-i\lambda_{+}h(z+x_{+})-i\lambda_{-}h(z+x_{-})-i\mu_{+}u(z+x_{+})
 \nonumber \\
&& -i\mu_{-}u(z+x_{-})-i\eta_{+}\xi(z+x_{+})-i\eta_{-}\xi(z+x_{-})),
\end{eqnarray}
and,

\begin{equation}
\Theta=\sum_{j}\theta\delta(z-y_{j}),
\end{equation}

then,   

\begin{eqnarray}
&&\rho W(h_{+},h_{-},u_{+},u_{-},\xi_{+},\xi_{-},x,t)=\cr \nonumber\\
&&\int\frac{d \lambda_{+}d \lambda_{-}d \mu_{+}d \mu_{-}d \eta_{+}d \eta_{-}}
{(2\pi)^{6}}e^{-i\lambda_{+}h_{+}-i\lambda_{-}h_{-}-i\mu_{+}u_{+}-i\mu_{-}u_{-}-i\eta_{+}
\xi_{+}-i\eta_{-}\xi_{-}}\langle\Theta\rangle.
\end{eqnarray}

We now derive equations for $\langle\Theta\rangle$ and $W$. Using the equations
(109-111)
we obtain,
\begin{eqnarray}
\langle\Theta\rangle_{t}=&-&i\lambda_{+}\langle (\frac{\alpha}{2}u_{+}^{2}+f_{+})\sum_{j}
\theta\delta(z-y_{j})\rangle \nonumber\\
&-&i\lambda_{-}\langle (\frac{\alpha}{2}u_{-}^{2}+f_{-})\sum_{j}
\theta\delta(z-y_{j})\rangle \nonumber\\
&-&i\mu_{+}\langle (-\alpha u_{+}u_{x_{+}}+f_{x_{+}})\sum_{j}\theta\delta(z-y_{j})\rangle 
 \nonumber\\
&-&i\mu_{-}\langle (-\alpha u_{-}u_{x_{-}}+f_{x_{-}})\sum_{j}\theta\delta(z-y_{j})\rangle
 \nonumber\\
&-&i\eta_{+}\langle (-\alpha \xi_{+}^{2}-\alpha u_{+}\xi_{x_{+}}+f_{x_{+}x_{+}})
\sum_{j}\theta\delta(z-y_{j})\rangle \nonumber\\ 
&-&i\eta_{-}\langle (-\alpha \xi_{-}^{2}-\alpha u_{-}\xi_{x_{-}}+f_{x_{-}x_{-}})
\sum_{j}\theta\delta(z-y_{j})\rangle \nonumber\\ 
&+&\langle\sum_{j}(-\alpha\bar{u}(y_{j},t))\delta^{1}(z-y_{j})\theta
+\sum_{k}(\delta(z-y_{j})\delta(t-t_{k})\theta\rangle\nonumber\\
&-&\langle\sum_{l}(\delta(z-y_{l})\delta(t-t_{l})\theta\rangle.
\end{eqnarray}

$\delta^{1}(z)=\frac{d}{dz}\delta(z)$ , the $(y_{k},t_{k})$'s are the
points of singularity creations and the $(y_{l},t_{l})$'s are the points of 
singularity annihilation due to collisions. Assuming homogeneity and using
the following identity [73],

\begin{equation}
\theta\delta^{1}(z-y_{j})=(\theta\delta(z-y_{j}))_{z}-\theta_{x_{+}}\delta
(z-y_{j})-\theta_{x_{-}}\delta(z-y_{j}),
\end{equation}

it follows that,
\begin{eqnarray}
\langle\Theta\rangle_{t}=&-&i\lambda_{+}\langle (\frac{\alpha}{2}u_{+}^{2}+f_{+})\sum_{j}
\theta\delta(z-y_{j})\rangle\nonumber\\
&-&i\lambda_{-}\langle (\frac{\alpha}{2}u_{-}^{2}+f_{-})\sum_{j}
\theta\delta(z-y_{j})\rangle \nonumber\\
&-&i\mu_{+}\langle (-\alpha u_{+}u_{x_{+}}+f_{x_{+}})\sum_{j}\theta\delta(z-y_{j})\rangle 
\nonumber\\
&-&i\mu_{-}\langle (-\alpha u_{-}u_{x_{-}}+f_{x_{-}})\sum_{j}\theta\delta(z-y_{j})\rangle
 \nonumber\\
&-&i\eta_{+}\langle (-\alpha \xi_{+}^{2}-\alpha u_{+}\xi_{x_{+}}+f_{x_{+}x_{+}})
\sum_{j}\theta\delta(z-y_{j})\rangle  \nonumber\\ 
&-&i\eta_{-}\langle (-\alpha \xi_{-}^{2}-\alpha u_{-}\xi_{x_{-}}+f_{x_{-}x_{-}})
\langle\sum_{j}{\bar u}(y_{j},t)(\theta_{x_{+}}+\theta_{x_{-}})\delta(z-y_{j})\rangle 
 \nonumber \\
&&+\Sigma_{1}-\Sigma_{2},
\end{eqnarray}

where $\Sigma_{1}$ and $\Sigma_{2}$ account respectively for singularity
creation and collision events. These are given by,

\begin{eqnarray}
&&\Sigma_{1}(\lambda_{+},\lambda_{-},\mu_{+},\mu_{-},\eta_{+},\eta_{-},x_{+},
x_{-},z,t)=\langle\sum_{k}\theta\delta(z-y_{k})\delta(t-t_{k})\rangle ,\\
&&\Sigma_{2}(\lambda_{+},\lambda_{-},\mu_{+},\mu_{-},\eta_{+},\eta_{-},x_{+},
x_{-},z,t)=\langle\sum_{l}\theta\delta(z-y_{l})\delta(t-t_{l})\rangle.
\end{eqnarray}

Invoking to the Novikov's theorem we have,

\begin{eqnarray}
\langle f_{\pm}\theta\rangle&=&(-i\lambda_{\pm} k(0)-i\lambda_{\mp} k(x_{\pm}-x_{\mp})-i\mu_
{\mp}k_{x}(x_{\pm}-x_{\mp}) \nonumber\\
&+&i\eta_{\pm}k_{xx}(0)+i\eta_{\mp}(x_{\pm}-x_{\mp}))
\langle\theta\rangle, \\
\langle f_{x_{\pm}}\theta\rangle&=&(-i\lambda_{\pm} k_{x}(x_{\pm}-x_{\mp})-i\mu_{\pm}k_{xx}(0)
+i\mu_{\mp}k_{x}(x_{\pm}-x_{\mp})\nonumber\\
&+&i\eta_{\mp}k_{xxx}(x_{\pm}-x_{\mp}))
\langle\theta\rangle, \\
\langle f_{x_{\pm}x_{\pm}}\theta\rangle&=&(-i\lambda_{\pm} k_{xx}(x_{\pm}-x_{\mp})\nonumber\\
&-&i\lambda_{\mp}k_{xx}(x_{\pm}-x_{\mp})
-i\mu_{\mp}k_{xxx}(x_{\pm}-x_{\mp})
+i\eta_{\pm}k_{xxxx}(0)\nonumber\\
&+&i\eta_{\mp}k_{xxxx}(x_{\pm}-x_{\mp}))\langle\theta\rangle.
\end{eqnarray}

To average the convective terms we use

\begin{eqnarray}
&&i\alpha\mu_{\pm}\langle u_{\pm}\xi_{\pm}\Theta\rangle+i\alpha\eta_{\pm}\langle 
u_{\pm}\xi_{\pm}
\Theta\rangle=-\alpha\langle
u_{\pm}\Theta_{x_{\pm}}\rangle+i\alpha\lambda_{\pm}\langle u^{2}_{\pm}
\Theta\rangle  \nonumber\\
&=&-\alpha\langle u_{\pm}\Theta\rangle_{x_{\pm}}+\alpha\langle\xi_{\pm}\Theta\rangle+i\alpha\lambda
_{\pm}\langle u_{\pm}^{2}\Theta\rangle  \nonumber \\
&=&-i\alpha\langle\Theta\rangle_{x_{\pm}\mu_{\pm}}+i\alpha\langle\Theta\rangle_{\eta_{\pm}}
-i\alpha\lambda_{\pm}\langle\Theta\rangle_{\mu_{\pm}\mu_{\pm}}.
\end{eqnarray}

Notifying that,
\begin{eqnarray}
\theta_{x_{\pm}}&=&(i\lambda_{\pm}u_{\pm}-i\mu_{\pm}\xi_{x_{\pm}}-i\eta_{\pm}
\xi_{x_{\pm}})\theta,\\
\langle u_{\pm}^{2}\Theta\rangle&=&-\langle\Theta\rangle_{\mu_{\pm}\mu_{\pm}},\\
\langle\xi_{\pm}^{2}\Theta\rangle&=&-\langle
\Theta\rangle_{\eta_{\pm}\eta_{\pm}},
\end{eqnarray}
 
Finally for $\langle\Theta\rangle_{t}$ we find,

\begin{eqnarray}
\langle\Theta\rangle_{t}&&=-\frac{i\alpha}{2}\lambda_{+}\langle\Theta\rangle_{\mu_{+}\mu_{+}}
-\frac{i\alpha}{2}\lambda_{-}\langle\Theta\rangle_{\mu_{-}\mu_{-}}\nonumber\\
&&-\frac{i\alpha}{2}(\langle\Theta\rangle_{x_{+}\mu_{+}}+\langle\Theta\rangle_{x_{-}\mu_{-}}\nonumber\\
&&-\langle\Theta\rangle_{x_{+}\mu_{-}}-\langle\Theta\rangle_{x_{-}\mu_{+}}+\frac{i\alpha}{2}
(\langle\Theta\rangle_{\eta_{+}}+\langle\Theta\rangle_{\eta_{-}})\nonumber\\
&&-i\alpha\eta_{+}
\langle\Theta\rangle_{\eta_{+}\eta_{+}}-i\alpha\eta_{-}\langle\Theta\rangle_{\eta_{-}\eta_{-}}\nonumber\\
&&-(\lambda_{+}^{2}k(0)+\lambda_{-}^{2}k(0)+2\lambda_{+}\lambda_{-}k(x_{+}-x_{-})
)\langle\Theta\rangle\nonumber\\
&&-(\mu_{+}^{2}k_{xx}(0)+\mu_{-}^{2}k_{xx}(0)+2\mu_{+}\mu_{-}k_{xx}(x_{+}-x_
{-}))\langle\Theta\rangle\nonumber\\
&&-(\eta_{+}^{2}k_{xxxx}(0)+\eta_{-}^{2}k_{xxxx}(0)+2\eta_{+}\eta_{-}k(x_{+}
-x_{-}))\langle\Theta\rangle\nonumber\\
&&-2(\lambda_{+}\mu_{-}-\lambda_{-}\mu_{+})k_{x}(x_{+}-x_{-})\langle\Theta\rangle
+2(\lambda_{+}\eta_{+}+\lambda_{-}\eta_{-})k_{xx}(0)\langle\Theta\rangle\nonumber\\
&&+2(\lambda_{+}\eta_{-}+\lambda_{-}\eta_{+})k_{xx}(x_{+}-x_{-})\langle\Theta\rangle
-2(\eta_{-}\mu_{+}-\eta_{+}\mu_{-})k_{xxx}(x_{+}-x_{-})\langle\Theta\rangle\nonumber\\
&&\langle\alpha\sum_{j}{\bar u}(y_{j},t)(\theta_{x_{+}}+\theta_{x_{-}})\delta(z-y_{j})\rangle 
+\Sigma_{1}-\Sigma_{2}.
\end{eqnarray}

for the term involving $\bar u(y_{j},t)$ we note that,

\begin{eqnarray}
&&u_{\pm}(y_{j},t)\theta_{x_{\pm}}=(u(y_{j}+x_{\pm},t)\theta)_{x_{\pm}}
-\xi(y_{j}+x_{\pm},t)\theta=i\theta_{x_{\pm}}\lambda_{\pm}-i\theta_{\mu_{\pm}},
\\
&&u_{\pm}(y_{j},t)\theta_{x_{\mp}}=(u(y_{j}+x_{\pm},t)\theta)_{x_{\mp}}
=i\theta_{x_{\mp}}\lambda_{\pm},
\end{eqnarray}

thus, 

\begin{eqnarray}
&&\alpha\langle\sum_{j}\bar u(y_{j},t)(\theta_{x_{+}}+\theta_{x_{-}})\delta(z-y_{j})
\rangle \nonumber\\
&&=\frac{i\alpha}{2}(\langle\Theta\rangle_{x_{+}\lambda_{+}}+\langle\Theta\rangle_{x_{-}\lambda_{+}}
+\langle\Theta\rangle_{x_{+}\lambda_{-}} \nonumber\\
&&+\langle\Theta\rangle_{x_{-}\lambda_{-}}-\langle\Theta\rangle_{\mu_{+}}
-\langle\Theta\rangle_{\mu_{-}}).
\end{eqnarray}

Combining the above expressions, on the subset
$\lambda_{+}=\lambda_{-}=\frac{\lambda}{2},x_{+}=-x_{-}=\frac{x}{2}
,\mu_{1}=\mu_{+}+\mu_{-},\mu_{2}=\frac{\mu_{+}-\mu_{-}}{2}$,
$\langle\Theta\rangle$ satisfies,

\begin{eqnarray}
&&\langle\Theta\rangle_{t}=-\frac{i\alpha}{4}\lambda(2\langle\Theta\rangle_{\mu_{1}\mu_{1}}
+\frac{1}{2}\langle\Theta\rangle_{\mu_{2}\mu_{2}})-{i\alpha}\langle\Theta\rangle_{x\mu_{2}}\nonumber\\
&&+\frac{i\alpha}{2}(\langle\Theta\rangle_{\eta_{+}}+\langle\Theta\rangle_{\eta_{-}})
-i\alpha\eta_{+}\langle\Theta\rangle_{\eta_{+}\eta_{+}}-
i\alpha\eta_{-}\langle\Theta\rangle_{\eta_{-}\eta_{-}}\nonumber\\
&&-\frac{\lambda^2}{2}(k(0)+k(x))\langle\Theta\rangle
-(\frac{\mu_{1}^2}{2}+2\mu_{2}^2)k_{xx}(0)-2(\frac{\mu_{1}^2}{4}-
\mu_{2}^2)k_{xx}(x)\langle\Theta\rangle\nonumber\\
&&-(\eta_{+}^2k_{xxxx}(0)+\eta_{-}^2k_{xxxx}(0)+2\eta_{+}\eta_{-}k_{xxxx}(x))
\langle\Theta\rangle+2\lambda\mu_{2}k_{x}(x)\langle\Theta\rangle\nonumber\\&&
+\lambda(\eta_{+}+\eta_{-})(k_{xx}(0)+k_{xx}(x))\langle\Theta\rangle
+2\mu_{2}(\eta_{+}+\eta_{-})k_{xxx}(x)\langle\Theta\rangle+\nonumber\\
&&\mu_{1}(\eta_{-}-\eta_{+})k_{xxx}(x)\langle
\Theta\rangle+\Sigma_{2}-\Sigma_{1}.\nonumber
\end{eqnarray}

The $\Sigma_{1},\Sigma_{2}$ are evaluated at 
$\lambda_{+}=\lambda_{-}=\frac{\lambda}{2},x_{+}=-x_{-}=\frac{x}{2}
,\mu_{1}=\mu_{+}+\mu_{-},\mu_{2}=\frac{\mu_{+}-\mu_{-}}{2}$.

Changing to the variables $(h_{v},{\bar u},s,\xi_{+},\xi_{-})$, we 
obtain the following equation for $W$,

\begin{eqnarray}{\label{w-eq1}}
&&(\rho W(h_{v},\bar u,s,\xi_{+},\xi_{-},x,t))_{t}=\frac{\alpha}{2}{\bar u}^2\rho
W_{h_{v}}+\frac{\alpha}{8}s^2\rho W_{h_{v}}-\alpha s \rho W_{x}\nonumber\\
&&+\frac{\alpha}{2}\xi_{+}\rho W +\frac{\alpha}{2}\xi_{-}\rho W 
+\alpha \rho ({\xi_{+}}^2 W)_{\xi_{+}}+\alpha \rho ({\xi_{-}}^2 W)_{\xi_{-}}
\nonumber\\
&&+k(0)\rho W_{h_{v}h_{v}}+k_{xx}(0)\rho W_{\bar u \bar u}
+2\rho(k_{xx}(0)-k_{xx}(x))W_{ss}\nonumber\\
&&+\rho k_{xxxx}(0)(W_{\xi_{+}\xi_{+}}+
W_{\xi_{-}\xi_{-}})+2\rho k_{xxxx}(x)W_{\xi_{-}\xi_{-}}-2k_{x}(x)\rho W_{sh_{v}}
\nonumber\\&&
-2k_{xx}(0)\rho W_{h_{v}\xi_{+}}-2k_{xx}(0)\rho W_{h_{v}\xi_{-}}
-2k_{xx}(x)\rho W_{h_{v}\xi_{+}}-2k_{xx}(x)\rho W_{h_{v}\xi_{-}}\nonumber\\
&&-2\rho k_{xxx}(x)(W_{s\xi_{-}}+W_{s\xi_{+}})
k_{xxx}(x)\rho W_{\bar u\xi_{-}}+k_{xxx}(x)\rho W_{\bar u\xi_{+}}\nonumber\\
&&+\zeta_{1}-\zeta_{2}.
\end{eqnarray}

The $\zeta_{1}(h_{v_{j}},{\bar u},s,\xi_{+},\xi_{-},x,t)$
is defined such that,  

\begin{equation}
\zeta_{1}(h_{v_{j}},{\bar u},s,\xi_{+},\xi_{-},x,t)d h_{v_{j}} ds d{\bar u}
d\xi_{+} d\xi_{-} dz dt,
\end{equation}

gives the average number of singularity creation points in 
$[z,z+dz)\times[t,t+dt)$ with,

\begin{eqnarray}
h_{v}(x,y_{1},t_{1})\in [h_{v},h_{v}+d h_{v}),\nonumber\\
{\bar u}(x,y_{1},t_{1})\in [{\bar u},{\bar u}+d \bar u),\nonumber\\
s(x,y_{1},t_{1})\in [s,s+d s),\nonumber\\
\xi(y_{1}+\frac{x}{2},t_{1})\in [\xi_{+},\xi_{+}+d \xi_{+}),\nonumber\\
\xi(y_{1}-\frac{x}{2},t_{1})\in [\xi_{-},\xi_{-}+d \xi_{-}),\nonumber
\end{eqnarray}

conditional on $(y_{1},t_{1})\in([z,z+d z)\times [t,t+dt))$ being a point of 
singularity creation (because of the statistical homogeneity, $z$ is a dummy 
variable).
$\zeta_{2}(h_{v_{j}},{\bar u},s,\xi_{+},\xi_{-},x,t)$
is defined such that,  

\begin{equation}
\zeta_{2}(h_{v_{j}},{\bar u},s,\xi_{+},\xi_{-},x,t)d h_{v_{j}} ds d{\bar u}
d\xi_{+} d\xi_{-} dz dt,
\end{equation}
gives the average number of singularity collision points in 
$[z,z+dz)\times[t,t+dt)$ with,

\begin{eqnarray}
h_{v}(x,y_{2},t_{2})\in [h_{v},h_{v}+d h_{v}),\nonumber\\
{\bar u}(x,y_{2},t_{2})\in [{\bar u},{\bar u}+d \bar u),\nonumber\\
s(x,y_{2},t_{2})\in [s,s+d s),\nonumber\\
\xi(y_{2}+\frac{x}{2},t_{2})\in [\xi_{+},\xi_{+}+d \xi_{+}),\nonumber\\
\xi(y_{2}-\frac{x}{2},t_{2})\in [\xi_{-},\xi_{-}+d \xi_{-}),\nonumber
\end{eqnarray}

conditional on $(y_{2},t_{2})\in([z,z+d z)\times [t,t+dt))$ being a point of 
singularity collision.
Now we rescale  $h_{v}$ as $h'_{v}=\frac{h_{v}}{L^{1/2}}$, so the 
equation (\ref{w-eq1})changes to, 

\begin{eqnarray}{\label{w-eq2}}
&&L^{-1/2}\frac{\alpha}{2}{\bar u}^2\rho
W'_{h'_{v}}+L^{-1/2}\frac{\alpha}{8}s^2\rho W'_{h'_{v}}-\alpha s \rho W'_{x}
+\frac{\alpha}{2}\xi_{+}\rho W' +\frac{\alpha}{2}\xi_{-}\rho W' \nonumber\\
&&+\alpha \rho ({\xi_{+}}^2 W')_{\xi_{+}}+\alpha \rho ({\xi_{-}}^2 W')_{\xi_{-}}
+L^{-1}k(0)\rho W'_{h'_{v}h'_{v}}+k_{xx}(0)\rho W'_{\bar u \bar u}\nonumber\\
&&+2\rho(k_{xx}(0)-k_{xx}(x))W'_{ss}+\rho k_{xxxx}(0)(W'_{\xi_{+}\xi_{+}}+
W'_{\xi_{-}\xi_{-}})\nonumber\\
&&+2\rho k_{xxxx}(x)W'_{\xi_{-}\xi_{-}}-2L^{-1/2}k_{x}(x)
\rho W'_{sh'_{v}}-
2k_{xx}(0)L^{-1/2}\rho W'_{h'_{v}\xi_{+}}\nonumber\\
&&-2k_{xx}(0)L^{-1/2}\rho W'_{h'_{v}\xi_{-}}
-2k_{xx}(x)L^{-1/2}\rho W'_{h'_{v}\xi_{+}}-2k_{xx}(x)L^{-1/2}\rho W'_{h'_{v}\xi_{-}}
\nonumber\\
&&-2\rho k_{xxx}(x)(W'_{s\xi_{-}}+W'_{s\xi_{+}})-
k_{xxx}(x)\rho W'_{\bar u\xi_{-}}+k_{xxx}(x)\rho W'_{\bar u\xi_{+}}\nonumber\\
&&+\zeta'_{1}-\zeta'_{2}=0.
\end{eqnarray}

In the limit of large $L$ or $L \rightarrow \infty$ the leading terms are,

\begin{eqnarray}{\label{w-eql}}
&&-\alpha s \rho W'_{x}
+\frac{\alpha}{2}\xi_{+}\rho W' +\frac{\alpha}{2}\xi_{-}\rho W' \nonumber\\
&&+\alpha \rho ({\xi_{+}}^2 W')_{\xi_{+}}+\alpha \rho ({\xi_{-}}^2 W')_{\xi_{-}}
+k_{xx}(0)\rho W'_{\bar u \bar u}\nonumber\\
&&+2\rho(k_{xx}(0)-k_{xx}(x))W'_{ss}+\rho k_{xxxx}(0)(W'_{\xi_{+}\xi_{+}}+
W'_{\xi_{-}\xi_{-}})\nonumber\\
&&+2\rho k_{xxxx}(x)W'_{\xi_{-}\xi_{-}}
-2\rho k_{xxx}(x)(W'_{s\xi_{-}}+W'_{s\xi_{+}})-\nonumber\\&&
k_{xxx}(x)\rho W'_{\bar u\xi_{-}}+k_{xxx}(x)\rho W'_{\bar u\xi_{+}}+
\zeta'_{1}-\zeta'_{2}=0.
\end{eqnarray}

To find the $g_{n,m}$ we multiply the above equation by
$h^{'n}{\bar u}^{m}s^{p}$ and integrating over $h',{\bar u},s,\xi_{+}$ and 
$\xi_{-}$ we have,

\begin{eqnarray}
&&-\alpha\rho\langle h^{'n}_{v} {\bar u}^{m}s^{p+1}\rangle_{x}
+\frac{\alpha\rho}{2}\langle h^{'n}_{v} {\bar u}^{m}s^{p}\xi_{+}\rangle
+\frac{\alpha\rho}{2}\langle 
h^{'n}_{v} {\bar u}^{m}s^{p}\xi_{-}\rangle\cr \nonumber\\
&&+2p(p-1)\rho[k_{xx}(0)-k_{xx}(x)]\langle 
h^{'n}_{v} {\bar u}^{m}s^{p-2}\rangle\cr \nonumber\\
&&+m(m-1)k_{xx}(0)\rho\langle h^{'n}_{v} {\bar u}^{m-2}s^{p}\rangle
+Q^{(1)}_{nmp}-Q^{(2)}_{nmp}=0,
\end{eqnarray}

where,

\begin{eqnarray}
Q^{(1)}_{nmp}&=&\int h^{'n}{\bar u}^{m}s^{p}\zeta'_{1} dh' d{\bar u} ds d\xi_{+}
d\xi_{-},\cr \nonumber\\
Q^{(2)}_{nmp}&=&\int h^{'n}{\bar u}^{m}s^{p}\zeta'_{2} dh' d{\bar u} ds d\xi_{+}
d\xi_{-}. \nonumber
\end{eqnarray}

Using the following identity,

\begin{eqnarray}
&&\frac{\partial}{\partial x}h'_{v}(x,y_{j},t)=-\frac{s}{4L^{1/2}},\\
&&\frac{\partial}{\partial x}s(x,y_{j},t)=\frac{1}{2}(\xi_{+}+\xi_{-}),\\
&&\frac{\partial}{\partial x}{\bar u}(x,y_{j},t)=\frac{s}{2},
\end{eqnarray}

we find,

\begin{eqnarray}
& &\frac{\alpha\rho}{2}\langle h_{v}^{'n}{\bar 
u}^{m}s^{p}(\xi_{+}+\xi_{-})\rangle =\frac{\alpha\rho}{p+1}
\Big\{\langle h_{v}^{'n}{\bar u}^{m}s^{p+1}\rangle_{x}\cr\nonumber\\
&+&\frac{n}{4L^{S/2}}\langle 
h_{v}^{'(n-1)}{S\bar u}^{m}s^{p+2}\rangle
+\frac{m}{2}\langle h_{v}^{,n}{\bar u}^{m-1}s^{p+2}\rangle\Big \}
\end{eqnarray}

So in the limit of $L\rightarrow \infty$ we have

\begin{eqnarray}{\label{gnm-eq}}
&&-\frac{p\alpha\rho}{p+1}\langle h^{'n}_{v} {\bar u}^{m}s^{p+1}\rangle_{x}
+\frac{m\alpha\rho}{2(p+1)}\langle 
h^{'n}_{v} {\bar u}^{m-1}s^{p+2}\rangle\cr \nonumber\\
&&+2p(p-1)\rho[k_{xx}(0)-k_{xx}(x)] 
\langle h_{v}^{'n}{\bar u}^{m}s^{p-2}\rangle\cr \nonumber\\
&&+m(m-1)k_{xx}(0)\rho\langle h_{v}^{'n} {\bar u}^{m-2}s^{p}\rangle
+Q^{(1)}_{nmp}-Q^{(2)}_{nmp}=0.
\end{eqnarray}

Assuming a stationary solution for the dynamical equation (\ref{w-eq1})
governed over $W$ and rescaling $h_{v}$ as $h'_{v}=\frac{h_{v}}{L^{1/2}}$
in the resulting differential equation we reach to eq.(\ref{w-eq2}).
Of course the mentioned equation {\it is} dependent on scale $L$ but 
being interested in the limit of $L\rightarrow \infty$ results to 
eq.(\ref{w-eql})
which is free of the explicit scale dependent terms in the leading order .
However we are faced with two very complicated terms namely $\zeta_1$ and 
$ \zeta_2$ which would be analysed . 
The origin of these terms are related to processes of sharp valley creation and 
annihilation. We argue that these processes are basically involving local 
interaction between nearby sharp valleys and 
effects of forcing, which its spatial 
correlation is assumed to be much less than system size, 
so they essentially would not carry any information about system size.
In this sense eq.(\ref{w-eql}) encodes the fact that the probability 
distribution $W$ is a scale invariant function of its argument 
$h'_{v}=\frac{h_{v}}{L^{1/2}}$ in the leading order.
The above property is deciphered in eq.(\ref{gnm-eq}) too but this time 
it is translated in terms of the scale independence of $g_{n,m}$s.\\ 
Also the equation for $W$ enables us to find the time evolution of the 
sharp valley characteristics.
For example multiplying the equation (\ref{w-eq1}) by $h_{v}$ and 
integrating over all variables we can derive the increasing rate of mean 
height of the singularities and noting that,

\begin{eqnarray}
&&\frac{\partial}{\partial x}h_{v}(x,y_{j},t)=-\frac{s}{4},\\
&&\frac{\partial}{\partial x}s(x,y_{j},t)=\frac{1}{2}(\xi_{+}+\xi_{-}),
\end{eqnarray}

we get,

\begin{equation}
\frac{d}{d t}\langle h_{v}\rangle(t)=-\frac{\alpha}{8}\langle 
4{\bar u}^{2}-s^2 \rangle.
\end{equation}

\section{conclusion}

We study the
problem of non-equilibrium surface growth described by forced KPZ 
equation in 1$+$1 dimensions. The forcing is a white in time Gauaaian 
noise but with a Gaussian correlation in space. Modelling a short range 
correlated noise we restrict our study to the case when the correlation 
length of the forcing is much smaller than the system size. In 
the non-stationary regime 
when the sharp valley structures are not yet developed we find an exact
form for the generating function of the joint fluctuations of height and
height gradient. We determine the time scale of the
sharp valley formation and the exact functional form of the 
time dependence in the height difference
moments at any given order. 
Investigating the stationary state we 
give a general expression of the mixed correlations of height and
height-gradient at any order, in terms of the quantities 
which characterise the sharp valley singular structures.
Through a careful analysis being done over the behaviour of the sharp valley 
environment,
we decipher the general finite size corrections to the scaling 
of an arbitrary $n$th moment, i.e. $\langle (h-\bar 
h)^n\rangle$, at any order. Recently Marinari etal. [23] have obtained
the corrections to the leading order
scaling in dimensions $D=2,3,4$,
in a high resolution simulation on the RSOS discrete model which is 
beleived to be in the universality class of the KPZ equation 
stirred with a white in time Gaussian noise and delta correlated in space. 
Hence they get,

\be{\label{parisi}}
w_{n}(L) \sim A_{n} L^{n\chi} (1+ B_{n} L^{-\omega}).
\ee

Irrespective of the dimension and moment order $n$, they observe the same 
sub-leading exponent $\omega$ always very close to unity (see also [91,92]).
 Through our 
calculations we succeed to obtain the finite size corrections 
analytically. 
However we have to remark that, due to working with finite correlated 
forcing a firm comparison between our results and numerical 
simulations is not possible. 
More precisely, in the present paper the limiting of $\nu\rightarrow 0$ is 
taken into account only when $\sigma$ is finite. 
Still the forcing correlation length is 
much smaller than the system size and height correlation length.
But the limiting of $\sigma\rightarrow 0$ is a singular limit in our 
calculations, and moreover, it is not a priori clear that the limits of 
$\nu\rightarrow 0$ and $\sigma\rightarrow 0$ commute at all.
However due to the scale independence of $g_{n,m}$'s, the eq.(\ref{h-mom2}) 
shows the general correction terms for $n$th order moment, all having the 
same sub-leading exponent $\omega=1/2$. The amplitudes $A_n$ and $B_n$ in 
eq.(\ref{parisi}) are given explicitly in terms of the functions 
$g_{n,m}$ defined on the sharp valley singularities.
The next step, left for the future, would be the calculation of 
$g_{n,m}$'s in terms of 
few known parameters, i.e. the forcing and diffusion coefficients. \\
Our analysis enables us to find the stochastic equations which
are governed over the dynamics of quantities characterising the cusp
singularities too. This translates the stationary non-equilibrium 
dynamics of the surface in terms of the dynamics of singularities 
in the stationary state. 
When the
system crosses over the time $t^*$, after which the first singularties are 
formed, it would be an important study to analyse the shape 
deformation of non-stationary height PDF $P(h',t)$ in time. 
We believe that the analysis followed in this paper is
quite suitable for the zero temperature limit in the problem of directed
polymer in the random potential with short range correlations [88].
The same method applied to KPZ equation in higher 
dimensions would be definitely one of the consequent goals of the present 
work. 
The main message which might be encoded in the present work is the 
importance of the statistical properties of the geometrical singular 
structures for understanding the strong 
coupling regime of Kardar-Parisi-Zhang equation in higher dimensions.\\

\vskip +1cm

{\bf Acknowledgement}\\

We would like to appreaciate J. Bec for kindly providing us some of his 
unpublished numerical results. We also thank C. Castellano for reading the 
manuscript and bringing our attention to some of the references and Roya
Mohayaee for helpful comments. We are indebted to Eric Vanden 
Eijnden for very informative discussions. \\

\appendix
\section
{\bf an alternative method for determining of the moments of height 
fluctuation before the formation of the singularities}

In this appendix we give the details of calculations of the scaling behaviour
of moments of height difference before the formation of singularities.
We know that the generating function $Z(\mu,\lambda,t)$ satisfies the 
following equation when $(\nu\rightarrow 0)$,

\begin{equation}
Z_{t}=i\gamma\lambda Z-\frac{i\lambda\alpha}{2} Z_{\mu\mu}-\lambda^2k(0)Z+
\frac{i\alpha\lambda}{\mu}Z_{\mu}+\mu^2k_{xx}(0)Z.
\end{equation} 

Let us write $Z(\mu,\lambda,t)$ as follow,

\begin{eqnarray}{\label{gen-exp1}}
&&Z(\mu , \lambda , t) := (1 + A(t)
\lambda ^{2} + C(t)\lambda ^{3} + F(t)
\lambda ^{4} + G(t)\lambda ^{5} + J(t)
\lambda ^{6} + M(t)\lambda ^{7} \nonumber\\&&+ B(t)\lambda \mu ^{2} 
+ D(t)\lambda ^{2}\mu ^{2} + E
(t)\lambda \mu ^{4} + H(t)\lambda ^{3}\mu ^{2}
 + K(t)\lambda ^{2}\mu ^{4} + L(t)\lambda
 ^{4}\mu ^{2} +\nonumber\\&& N(t)\lambda ^{5}\mu ^{2}  
 + P(t)\lambda ^{3}\mu ^{4} + Q
(t)\lambda \mu ^{6})\exp{(( - \lambda ^{2}k(0) + \mu
 ^{2}{k_{{xx}}}(0))t)}.
\end{eqnarray}

Now expanding $Z(\mu,\lambda,t)$ as a series of $\mu,\lambda$
and substituting it in the equation eq.(\ref{gen-eq3}), we equate the 
terms in different orders of  
$\mu,\lambda$ ending with some coupled differential equations governed 
over the coefficients introduced in the definition of $Z$ in 
eq.(\ref{gen-eq3}). So we have,

\begin{eqnarray}
&&{\frac {\partial }{\partial t}}A(t)=i\alpha B(t), \\ &&
{\frac {\partial }{\partial t}}B(t)= - 2i\alpha 
{k_{{xx}}}(0)^{2}t^{2}, 
\\&&
{\frac {\partial }{\partial t}}C(t)= i\alpha D(t),
\\&&
{\frac {\partial }{\partial t}}D(t)= - 4i\alpha 
{k_{{xx}}}(0)tB(t) - 2i\alpha {E
}(t), 
\\&&
{\frac {\partial }{\partial t}}E(t)=0 ,
\\&&
{\frac {\partial }{\partial t}}F(t)=i\alpha 
H(t),  
\\&&
{\frac {\partial }{\partial t}}H(t)= - 4i\alpha 
{k_{{xx}}}(0)tD(t) - 2i\alpha {K
}(t) - 2i\alpha {k_{{xx}}}(0)^{2}t^{2}A
(t),   
\\&&
{\frac {\partial }{\partial t}}K(t)= - 2i\alpha 
{k_{{xx}}}(0)^{2}t^{2}B(t) - 8i\alpha {
k_{{xx}}}(0)tE(t) - 9i\alpha Q
(t),   
\\&&
{\frac {\partial }{\partial t}}Q(t)=0 ,
\end{eqnarray}

By solving these differential equations with the initial conditions 
that $A(t),B(t),C(t),D(t),E(t),F(t),H(t),K(t),Q(t)$ are zero at $t=0$,
we find, 

\begin{eqnarray}
&&A(t)=\frac{1}{6}\alpha^2k_{xx}(0)^2t^4,
\\&&
B(t)=-\frac{2}{3}i\alpha k_{xx}(0)^2t^3,
\\&&
C(t)=\frac{4}{45}i\alpha^3 k_{xx}(0)^3t^6,
\\&&
D(t)=\frac{8}{15}\alpha^2 k_{xx}(0)^3t^5,
\\&&
E(t)=0,
\\&&
F(t)=-\frac{101}{2520}\alpha^4 k_{xx}^4t^8,
\\&&
H(t)=\frac{101}{315}i\alpha^3 k_{xx}^4t^7,
\\&&
K(t)=-\frac{2}{9}\alpha^2 k_{xx}(0)^4t^6,
\\&&
Q(t)=0.
\end{eqnarray}

By replacing these expressions in eq.(\ref{gen-exp1}) we find 
$Z(\mu,\lambda,t)$ 
as a function of $\mu,\lambda,t$ explicitly without any unknown terms or 
expressions.
Now if we expand the original form of generating function $Z(\mu,\lambda,t)$
as a series in $\mu,\lambda$ we find,

\begin{eqnarray}{\label{gen-exp2}}
&&Z(\mu,\lambda,t)=\langle\exp{i\lambda((h-\bar h))+i\mu(\partial_{x} (h-\bar h))}\rangle= \nonumber\\
&& - { \frac {1}{720}} u^{6}\mu ^{6} - 
{ \frac {1}{120}} (h-\bar h)u^{5}\mu ^{5}\lambda  - 
{ \frac {1}{48}} (h-\bar h)^{2}u^{4}\mu ^{4}\lambda 
^{2} - { \frac {1}{36}} (h-\bar h)^{3}u^{3}\mu ^{3}
\lambda ^{3} \nonumber\\
&&- { \frac {1}{48}} (h-\bar h)^{4}u^{2}\mu 
^{2}\lambda ^{4}
 - { \frac {1}{120}} (h-\bar h)^{5}u\mu 
\lambda ^{5} - { \frac {1}{720}} (h-\bar h)^{6}\lambda ^{
6} - { \frac {1}{120}} iu^{5}\mu ^{5}\nonumber\\ 
&&-{ \frac {1}{24}} i(h-\bar h)u^{4}\mu ^{4}\lambda 
 - { \frac {1}{12}} i(h-\bar h)^{2}u^{3}\mu ^{3}
\lambda ^{2}
  - { \frac {1}{12}} i(h-\bar h)^{3}u^{2}
\mu ^{2}\lambda ^{3} - { \frac {1}{24}} i(h-\bar h)^{4}
u\mu \lambda ^{4} \nonumber\\
&&- { \frac {1}{120}} i(h-\bar h)^{
5}\lambda ^{5} + { \frac {1}{24}} u^{4}\mu ^{4
} + { \frac {1}{6}} (h-\bar h)u^{3}\mu ^{3}\lambda 
  + { \frac {1}{4}} (h-\bar h)^{2}u^{2}\mu 
^{2}\lambda ^{2} \nonumber\\
&&+ { \frac {1}{6}} (h-\bar h)^{3}u\mu
 \lambda ^{3} + { \frac {1}{24}} (h-\bar h)^{4}\lambda 
^{4} + { \frac {1}{6}} iu^{3}\mu^3+
{ \frac {1}{2}} i(h-\bar h)u^{2}\mu ^{2}\lambda   \nonumber\\&&
+{ \frac {1}{2}} i(h-\bar h)^{2}u\mu \lambda ^{2}  
 + { \frac {1}{6}} i(h-\bar h)^{3}\lambda 
^{3} - { \frac {1}{2}} u^{2}\mu ^{2} - (h-\bar h)u\mu
 \lambda \nonumber\\ 
&& - { \frac {1}{2}} (h-\bar h)^{2}\lambda ^{2}
 - iu\mu  - i(h-\bar h)\lambda  + 1.
\end{eqnarray}

Equating the coefficients of eq.(\ref{gen-exp1}) and eq.(\ref{gen-exp2}) 
proportional to 
the same powers in $\mu$ and $\lambda$ and replacing the expressions of 
$A(t),B(t),C(t), D(t),E(t),F(t),H(t),K(t),Q(t)$ it results to the same 
expressions given before, i.e. 

\begin{eqnarray}
&&\langle (h-\bar{h})^2\rangle=-\frac {1}{3}t(k_{xx}(0)^{2}\alpha 
^{2}t^{3}-6{k}(0)), \\
&&\langle (h-\bar{h})^3\rangle=-\frac {24}{45}k_{xx}(0)^{3}\alpha ^{3}t 
^{6},\\ 
&&\langle (h-\bar{h})^4\rangle=-\frac{101}{105}k_{xx}(0)^{4}\alpha
^{4}{t}^{8}-\frac{1}{6}t^{5}k_{xx}(0)^{2} \alpha ^{2}{k}(0)
+\frac{1}{2}t^{2}
k(0)^{2}.
\end{eqnarray}

\section
{\bf The proof of relation between $\rho$ and $\langle s^3 \rangle$}

We consider the statistical steady 
state i.e. $ R_t =0$ so that the eq.(\ref{req1}) can be written as 
follows,

\be{\label{req2}}
R_{uu}=\frac{1}{k_{xx}(0)}G(u,t)=\frac{\alpha \rho}{k_{xx}(0)}
(\int_{-\infty}^{0} ds\int_{u+\frac{s}{2}}
^{u-\frac{s}{2}}d{\bar u}(u-{\bar u})S({\bar u,s,t}))_{u}.
\ee
We integrate the eq.(\ref{req2}) w.r.t $u$ and find,
\be
R_{u}=\frac{ \alpha \rho}{k_{xx}(0)}
\int_{-\infty}^{0} ds\int_{u+\frac{s}{2}}
^{u-\frac{s}{2}}d{\bar u}(u-{\bar u})S({\bar u,s,t}).
\ee
At the large time limit ($t \rightarrow \infty$) we denote $R$ and $S$ as
$R_{\infty}$ and $S_{\infty} (\bar u,s)$. So, 
\be
R_{\infty}=\frac{\alpha \rho}{k_{xx}(0)}
\int_{-\infty}^{u} du\int_{-\infty}^{0} ds\int_{u+\frac{s}{2}}
^{u-\frac{s}{2}}d{\bar u}(u-{\bar u})S_{\infty}({\bar u,s}).
\ee
To determine the $R_{\infty}$ we define function $K(u)$ as the following,
\be{\label{k-eq1}}
K(u)=\frac{ \alpha\rho}{k_{xx}(0)}
\int_{-\infty}^{0} ds\int_{u+\frac{s}{2}}
^{u-\frac{s}{2}}d{\bar u}\frac{(u-{\bar u})^2}{2}S_{\infty}({\bar u,s}).
\ee
Differentiation the above equation with respect to $u$ gives us,
\bea{\label{k-eq2}}
&&\frac{d}{du}K(u)=\frac{ \alpha \rho}{k_{xx}(0)}
\int_{-\infty}^{0} ds\int_{u+\frac{s}{2}}
^{u-\frac{s}{2}}d{\bar u}(u-{\bar u})S_{\infty}({\bar u,s}) \cr \nonumber \\
&&\frac{ \alpha \rho}{k_{xx}(0)}\int_{-\infty}^{0} ds\frac{s^2}{8}
S_{\infty}({u-\frac{s}{2}},s)-\frac{ \alpha \rho}{k_{xx}(0)}
\int_{-\infty}^{0} ds\frac{s^2}{8}S_{\infty}({u+\frac{s}{2}},s)  .
\eea
Now we integrate the eq.(\ref{k-eq2}) over $u$ from $ - \infty $ to $u$ and 
find, 
\bea
&&\int_{-\infty}^{u} du \frac{d}{du}K(u)=\frac{ \alpha 
\rho}{k_{xx}(0)}\int_{-\infty}^{u} du
\int_{-\infty}^{0} ds\int_{u+\frac{s}{2}}
^{u-\frac{s}{2}}d{\bar u}(u-{\bar u})S_{\infty}({\bar u,s}) 
\cr \nonumber \\ 
&&\frac{ \alpha  \rho}{2k_{xx}(0)}\int_{-\infty}^{u} du
\int_{-\infty}^{0} ds\frac{s^2}{4}S_{\infty}({u-\frac{s}{2}},s)  
-\frac{  \alpha   \rho}{2k_{xx}(0)}\int_{-\infty}^{u} du
\int_{-\infty}^{0} ds\frac{s^2}{4}S_{\infty}({u+\frac{s}{2}},s).  
\eea
Then we will find,
\bea
K(u)-K(-\infty)&&=R_{\infty}(u)
+\frac{  \alpha   \rho}{2k_{xx}(0)}\int_{-\infty}^{0} ds
\int_{-\infty}^{u-\frac{s}{2}} d\bar u\frac{s^2}{4}
S_{\infty}(\bar u,s) \cr \nonumber \\ 
&&-\frac{   \alpha  \rho}{2k_{xx}(0)}
\int_{-\infty}^{0} ds\int_{-\infty}^{u+\frac{s}{2}} 
d\bar u\frac{s^2}{4}S_{\infty}(\bar u,s).  
\eea
According to the definition of $K(u)$ we see that $ K( - \infty) 
\rightarrow 0$ 
( the shock probability density function goes to zero in this limit) 
and therefore we find the following relation between 
$K(u)$ and $R_{\infty} (u)$,
\be{\label{k-eq3}}
K(u)=R_{\infty}(u)
+\frac{   \alpha  \rho}{2k_{xx}(0)}
\int_{-\infty}^{0} ds\int_{u+\frac{s}{2}}^{u-\frac{s}{2}} 
d\bar u\frac{s^2}{4}S_{\infty}(\bar u,s).  
\ee
Using the eqs.(\ref{k-eq1}) and (\ref{k-eq3}) we find explicit relation 
between the $R_{\infty }$ and $S_{\infty} (\bar u,s)$ as follows,
\be{\label{req3}}
R_{\infty}(u)=
-\frac{ \alpha  \rho}{2k_{xx}(0)}\int_{-\infty}^{0} ds
\int_{u+\frac{s}{2}}^{u-\frac{s}{2}} d\bar u 
[\frac{s^2}{4}-(u-\bar u)^2]S_{\infty}(\bar u,s).  
\ee
Assuming $S_{\infty} (\bar u, s) \geq 0$
, it becomes evident that above integral would give a realisable 
portability density for height gradient, that is $R_{\infty} \geq 0$.
For finite $\sigma$ the eq.(\ref{req3}) gives us the PDF of height gradient
in the KPZ equation in the strong coupling limit.
The function $ R_{\infty}(u))$ enables us to determine the relation between
the vallies density $\rho$ and $k_{xx}(0)$. We would integrate over $u$ 
from $R_{\infty}$ so we define another function $K_1 (u)$ such that,
\be
K_1(u)=
-\frac{ \alpha  \rho}{2k_{xx}(0)}\int_{-\infty}^{0} ds\int_{u+\frac{s}{2}}^{u-\frac{s}{2}} 
d\bar u [\frac{s^2}{4}u-\frac{(u-\bar u)^3}{3}]S_{\infty}(\bar u,s),  
\ee
where differentiation $K_1 (u)$ with respect to $u$ gives, 
\bea
\frac{d}{du}K_1 (u)&&=-\frac{\alpha  \rho}{2k_{xx}(0)}\int_{-\infty}^{0} ds\int_{u+\frac{s}{2}}^
{u-\frac{s}{2}} d\bar u [\frac{s^2}{4}-(u-\bar u)^2]S_{\infty}(\bar u,s) \cr \nonumber \\ 
&&-\frac{\alpha  \rho}{2k_{xx}(0)}\int_{-\infty}^{0} ds [\frac{s^2}{4}u-\frac{s^3}{24}]
S_{\infty}(u-\frac{s}{2},s) \cr \nonumber \\
&&+\frac{\alpha  \rho}{2k_{xx}(0)}\int_{-\infty}^{0} ds 
[\frac{s^2}{4}u+\frac{s^3}{24}]S_{\infty}(u+\frac{s}{2},s).
\eea
Now integrating the above equation over $u$ from $-\infty$ to 
$+\infty$ gives,
\bea
K_1(+\infty)-K_1(-\infty) &&= 
\int_{-\infty}^{+\infty} du R_{\infty}(u)\cr \nonumber\\
&&-\frac{\alpha  \rho}{2k_{xx}(0)}
\int_{-\infty}^{+\infty} du
\int_{-\infty}^{0} ds [\frac{s^2}{4}(u+\frac{s}{2})-\frac{s^3}{24}]
S_{\infty}(u,s) \cr \nonumber \\
&&+\frac{\alpha  \rho}{2k_{xx}(0)}\int_{-\infty}^{+\infty} 
du\int_{-\infty}^{0} ds 
[\frac{s^2}{4}(u-\frac{s}{2})+\frac{s^3}{24}]S_{\infty}(u,s).  
\eea

Using the fact that $  K_1(+\infty) = K_1(-\infty)= 0$, 
we obtain,
\bea
\int_{-\infty}^{+\infty} du R_{\infty}(u)&& = 
-\frac{\alpha \rho}{2k_{xx}(0)}\int_{-\infty}^{+\infty} du\int_{-\infty}^{0} ds 
[\frac{s^2}{4}u - \frac{s^3}{12}]  S_{\infty}(u,s) \cr \nonumber \\
&&+\frac{\alpha \rho}{2k_{xx}(0)}
\int_{-\infty}^{+\infty} du\int_{-\infty}^{0} ds 
[\frac{s^2}{4}u+\frac{s^3}{12}]S_{\infty}(u,s),  
\eea
in which the sum of the terms in right hand side gives,
\be
\int_{-\infty}^{+\infty} du R_{\infty}(u) =
\frac{ \alpha \rho}{12k_{xx}(0)}\langle s^3\rangle.
\ee
Thus from the requirement that $R_\infty$ be normalised to unity, we get,
\be{\label{kxx}}
k_{xx}(0)=\frac{\alpha \rho}{12}\langle s^3 \rangle.
\ee

\section
{\bf Derivation of the finite contribution of relaxation term
in the stationary state}
In this appendix we give the detail of calculations of 
$g_{n,m}$ in the equation eq.(\ref{gnm}).
For computing the $g_{nm}$ we introduce, 

\be
K(u)=-m(m-1)\alpha\rho\int_{-\infty}^{0} ds \int_{-\infty}^{\infty} dh' h'^{n}
\int_{u-\frac{s}{2}}^{u+\frac{s}{2}}  d \bar u 
\{ \frac{u^{m-1}}{m-1}{\bar u} - \frac{u^m}{m} \} S({\bar u},s, h',t).
\ee

By differentiating $K(u)$ and integrating in the whole range of $u$ 
we have,

\bea
& &\int_{-\infty}^{\infty} \frac{dK(u)}{du}du=g_{nm}\cr \nonumber \\
& &-m(m-1)\alpha\rho\int_{-\infty}^{0} ds \int_{-\infty}^{\infty} dh' h'^{n}
\int_{-\infty}^{\infty} du 
\{ \frac{u^m}{m}-\frac{u^{m-1}}{m-1}(u-\frac{s}{2}) \}
S(u-\frac{s}{2},s, h',t)\cr \nonumber \\
& &-m(m-1)\alpha\rho\int_{-\infty}^{0} ds \int_{-\infty}^{\infty} dh' h'^{n}
\int_{-\infty}^{\infty} du \{\frac{u^m}{m}-\frac{u^{m-1}}{m-1}(u+\frac{s}{2})\}
S(u+\frac{s}{2},s, h',t).
\eea

Since $K(+\infty)=K(-\infty)=0$, left hand side vanishes, so,  

\bea
g_{nm}&=&m(m-1)\alpha\rho\int_{-\infty}^{0} ds \int_{-\infty}^{\infty} dh' 
h'^{n}
\int_{-\infty}^{\infty} d{\bar u}[\frac{({\bar u}+\frac{s}{2})^m}{m}
-\frac{({\bar u}+\frac{s}{2})^{m-1}}{m-1}{\bar u}]S({\bar u},s, h',t) \cr \nonumber \\
&-&m(m-1)\alpha\rho\int_{-\infty}^{0} ds \int_{-\infty}^{\infty} dh' h'^{n}
\int_{-\infty}^{\infty} d{\bar u}[\frac{({\bar u}-\frac{s}{2})^m}{m}
-\frac{({\bar u}-\frac{s}{2})^{m-1}}{m-1}{\bar u}]S({\bar u},s, h',t) \cr \nonumber \\
&=&\frac{\rho}{2^m}\int_{-\infty}^{0} ds \int_{-\infty}^{\infty} d{\bar u}
\int_{-\infty}^{\infty} dh' h'^{n}\{({2\bar u}+s)^{m-1}\cr \nonumber \\
&&\{[(m-1)s-2{\bar u}]+(2{\bar u}-s)^{m-1}[(m-1)s+2{\bar u}] \}S({\bar 
u},s, h',t), 
\eea

which finally leads to eq.(\ref{gnm}).

\section
{\bf Dynamics of quantities which are defined on the sharp valleys}
In this section we determine the equation of motion for 
 ${\bar u}(y_j,t) , s(y_j,t) , {\tilde h}(y_j,t)$ along the sharp valley  
 which is located at position  $ y_j $ at time $t$.
 Using  the KPZ equation and 
its differentiation by $x$ around the sharp valley at position
 $ y_j$, one can find a set of equation for 
$h_+(y_j,t) = \lim_{\epsilon \rightarrow 0^+} h (y_j + \epsilon ,t)$,
$h_-(y_j,t) = \lim_{\epsilon \rightarrow 0^+} h (y_j - \epsilon ,t)$,
$u_+(y_j,t) = \lim_{\epsilon \rightarrow 0^+} u (y_j + \epsilon ,t)$, and
$u_-(y_j,t) = \lim_{\epsilon \rightarrow 0^+} u (y_j - \epsilon ,t)$ 
as follows,

\bea
{h_{+}}_t(y_j,t)&=&\frac{\alpha}{2}u_{+}^{2}(y_j,t)+f(y_j,t),\\
{h_{-}}_t(y_j,t)&=&\frac{\alpha}{2}u_{-}^{2}(y_j,t)+f(y_j,t),\\
{u_{+}}_t(y_j,t)&=&-\alpha u_{+}(y_j,t){u_{+}}_x(y_j,t)-f_x(y_j,t),\\
{u_{-}}_t(y_j,t)&=&-\alpha u_{-}(y_j,t){u_{-}}_x(y_j,t)-f_x(y_j,t).
\eea

To determine the $\frac{d}{dt} \{ \bar u , s, \tilde h \}$ we use
the following identity [73,84], 
\bea
\frac{d}{dt}u_{+}(y_j,t)&=&\frac{dy_j}{dt}{u_{+}}_x(y_j,t)
+{u_{+}}_t(y_j,t) \nonumber \\
&=&\alpha \bar u (y_j,t){u_+}_x(y_j,t)- \alpha { u_{+}}_x (y_j,t) {u_{+} (y_j,t)}
- f_x(y_j,t)  \nonumber \\
&=& - \frac{\alpha}{2}s(y_j,t){u_+}_x(y_j,t) - f_x(y_j,t),
\eea

where ${\bar u}=\frac{1}{\alpha}\frac{dy_j}{dt}$. Similarly,

\be
\frac{d}{dt}{\bar u}(y_{j},t)=\frac{\alpha}{2}s(y_{j},t){u_{-}}_{x}(y_{j},t)-
f_{x}(y_{j},t).
\ee

These equations can be re-written as, 

\bea
\frac{d}{dt}{\bar u}(y_j,t)&=&-\frac{\alpha}{4}s({u_+}_x-{u_-}_x)-f_x, 
\cr \nonumber \\
\frac{d}{dt}{s}(y_j,t)&=&-\frac{\alpha}{2}s({u_+}_x+{u_-}_x),
\eea
where will give the equations for $\bar u$ and $s$. Since
 $u=-h_x$ we write the above equations in term of curvature of the 
 surface in the right and left sides of the sharp valley  
at position $y_j$ as, 

 \bea
\frac{d}{dt}{\bar 
u}(y_j,t)&=&\frac{\alpha}{4}s({h_+}_{xx}-{h_-}_{xx})-f_x, \cr \nonumber \\
\frac{d}{dt}{s}(y_j,t)&=&\frac{\alpha}{2}s({h_+}_{xx}+{h_-}_{xx}).
\eea

For determining the time evolution of the $\tilde h = h - \bar h$, we use 
the KPZ equation by which one can easily show that $h_+(y_j,t)$  and $ 
h_-(y_j,t) $ satisfy,

\bea
\frac{d}{dt}h_+(y_j,t)&=&\frac{dy_j}{dt}{h_{+}}_x(y_j,t)+{h_{+}}_t(y_j,t), \cr \nonumber \\
\frac{d}{dt}h_-(y_j,t)&=&\frac{dy_j}{dt}{h_{-}}_x(y_j,t)+{h_{-}}_t(y_j,t).
\eea

By definition we have $\frac{d}{dt}h(y_j,t)=\frac{1}{2}(\frac{d}{dt}h_+ + 
\frac{d}{dt}h_-)$, so using the equation for $h_+$ and $h_-$,  

\be
\frac{d}{dt}h(y_j,t) =-\frac{\alpha}{8}(4{\bar u}^2-s^2) + f,
\ee
and 

\be
\frac{d}{dt}{\tilde h}(y_j,t) =- \frac{\alpha}{8}(4{\bar u}^2-s^2) + 
f - \gamma,
\ee

where ${\tilde h}(y_j,t)=h(y_j,t)-{\bar h}$ and ${\bar h}_t=\gamma$.

Therefore in summary we have the following set of equations for given sharp valley  
in the KPZ problem in the limit $\nu\rightarrow 0$,
\bea 
\frac{dy_j}{dt} &=& \alpha \bar u,  \cr \nonumber \\
\frac{d}{dt}{\bar 
u}(y_j,t)&=&\frac{\alpha}{4}s({h_+}_{xx}-{h_-}_{xx})-f_x, \cr \nonumber \\
\frac{d}{dt}{s}(y_j,t)&=&\frac{\alpha}{2}s({h_+}_{xx}+{h_-}_{xx}), \cr 
\nonumber \\
\frac{d}{dt}{\tilde h}(y_j,t) &=&-\frac{\alpha}{8}(4{\bar u}^2-s^2) + f 
- \gamma.
\eea

\newpage
\vspace{ 4cm}
\begin{center}
{\bf  Figure Captions}
\end {center}

\vspace{4cm}
{\bf Figure.1} { Behaviour of the time 
scales in which the moments $\langle (h-{\tilde h})^{2n}\rangle$ become 
negative in terms of $n$. The square points 
are calculated according to eq.(\ref{gen-s2}) 
while the solid line is the fitting curve asymptotically tending to 
$\frac{1}{4}(\frac{k(0)}{\alpha^2 k_{xx}^2(0)})^{1/3}$.} \\

{\bf Figure.2} { In the upper graph 
the sharp valley solutions in KPZ equation are demonstrated while in lower one 
the corresponding shock structures in Burgers equation are sketched . The 
variables characterising the cusp, namely $h_{x-}$ , $h_{x+}$ and 
${\tilde h}$ are shown.\\

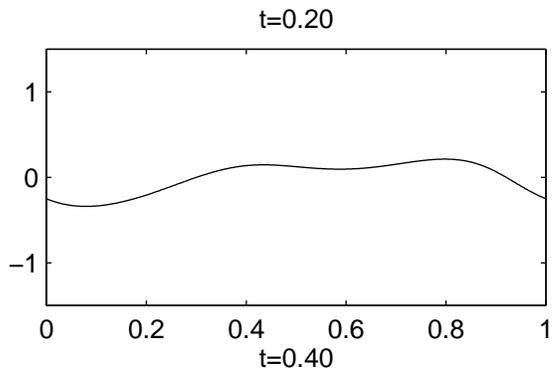
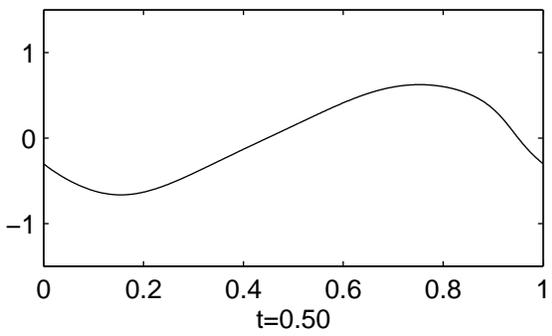
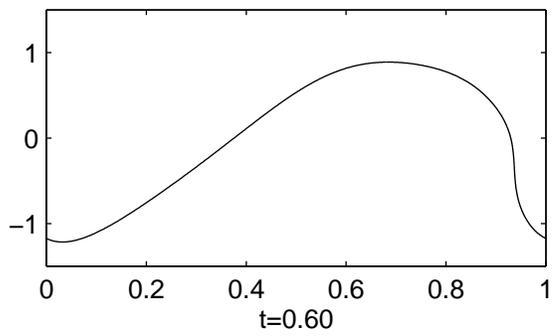
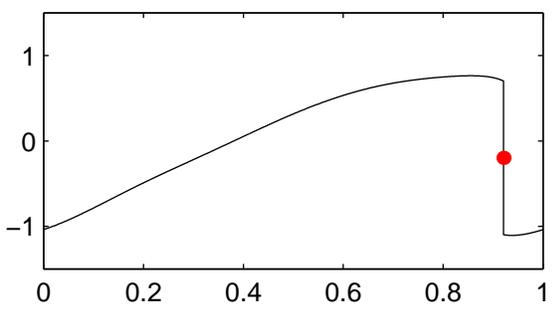
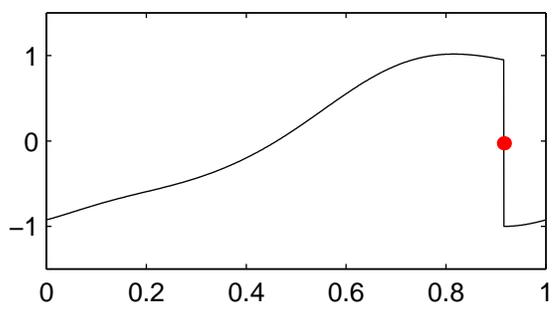
{\bf Figure.3} { 
Different time snapshots of gradient configuration
whithin system size, i.e. $-\partial_x h$ verses $x$. The time
scale for shock creation is demonstrated for $\sigma \sim L$. 
The solid points are showing the 
jumps in the height gradient. {\it J. Bec} [93] }. \\

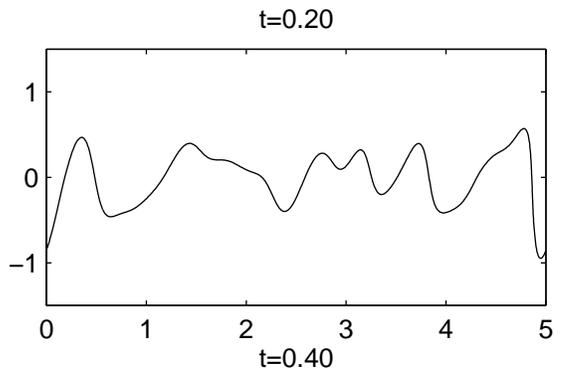
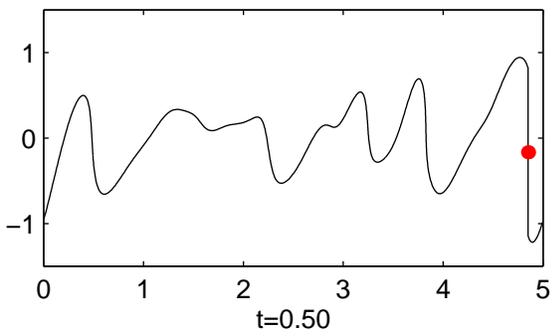
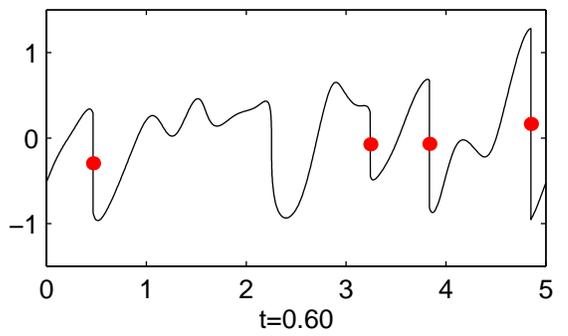
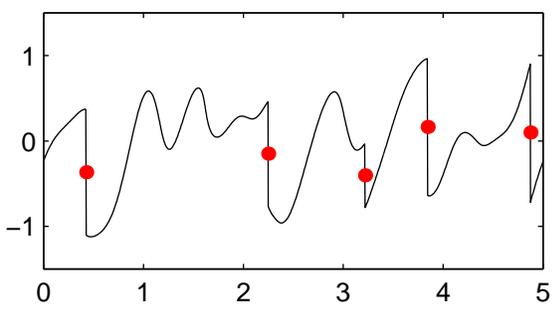
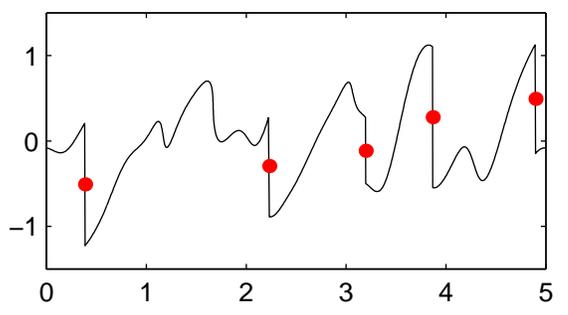
{\bf Figure.4} {Different time snapshots of gradient configuration 
whithin system size, i.e. $-\partial_x h$ verses $x$. The time 
scale for shock creation 
is demonstrated for $\sigma <  L$. The solid points are showing the
jumps in the height gradient. {\it J. Bec }[93] }. \\

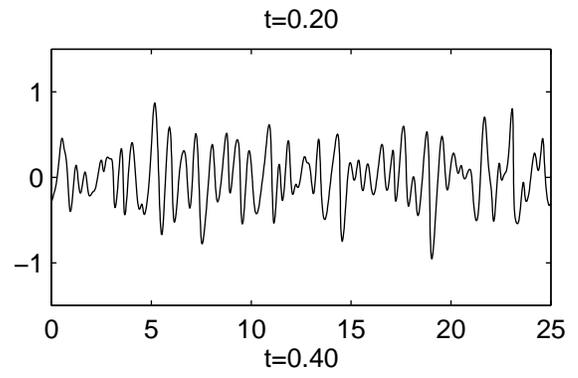
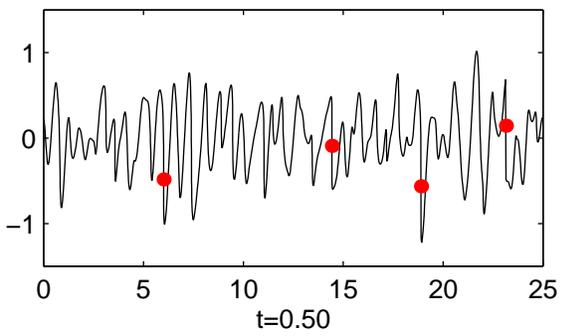
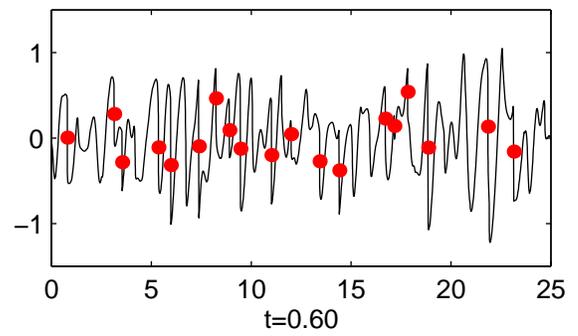
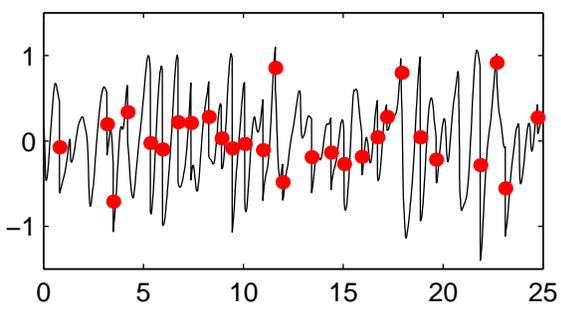
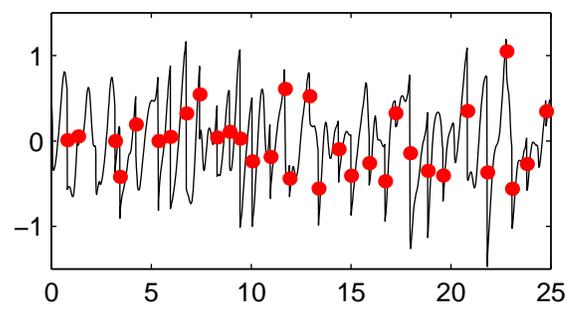
{\bf Figure.5} { 
Different time snapshots of gradient configuration
whithin system size, i.e. $-\partial_x h$ verses $x$. The time
scale for shock creation
is demonstrated for $\sigma \ll  L$. The solid points are showing the
jumps in the height gradient. {\it J. Bec }[93].
}

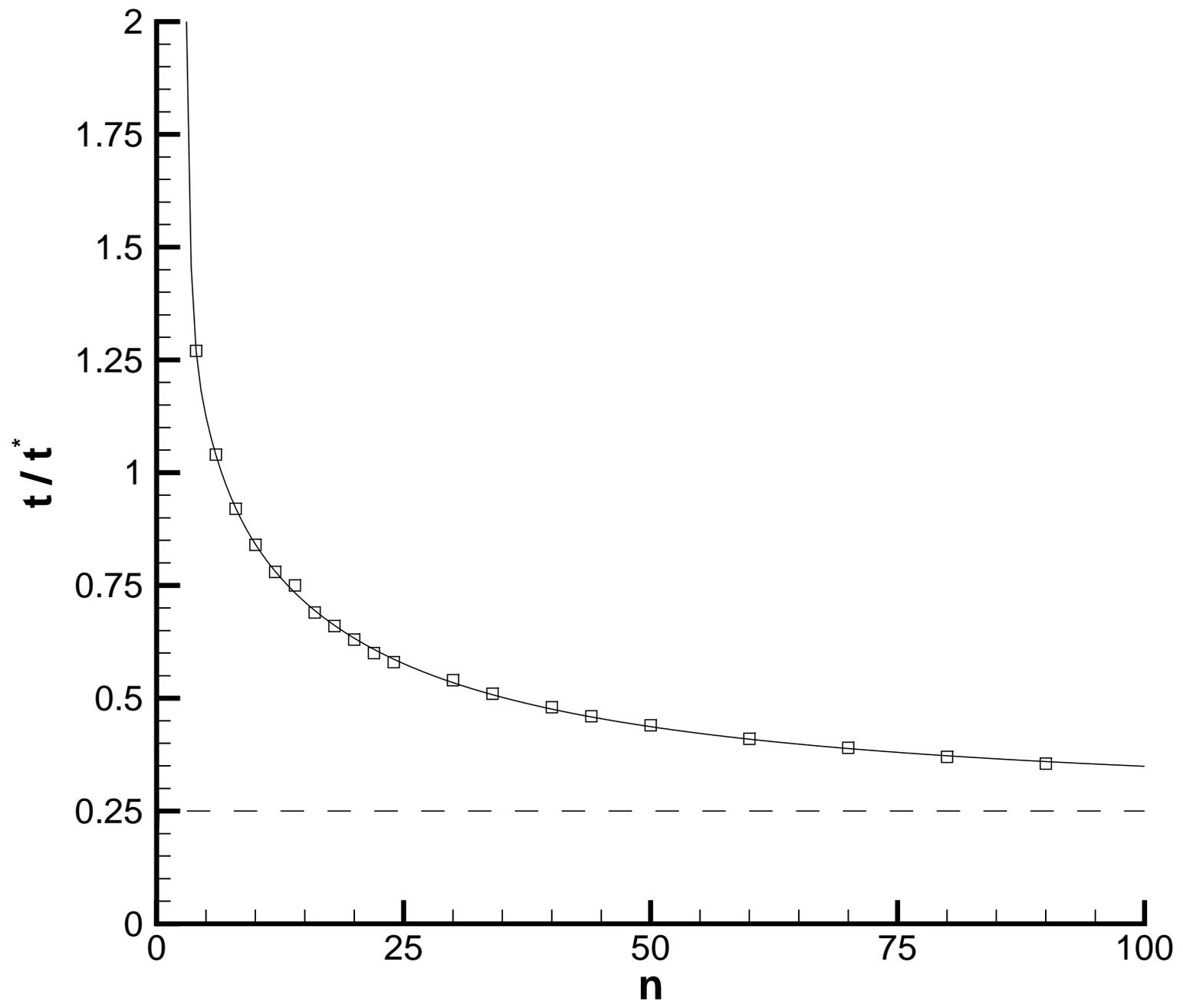
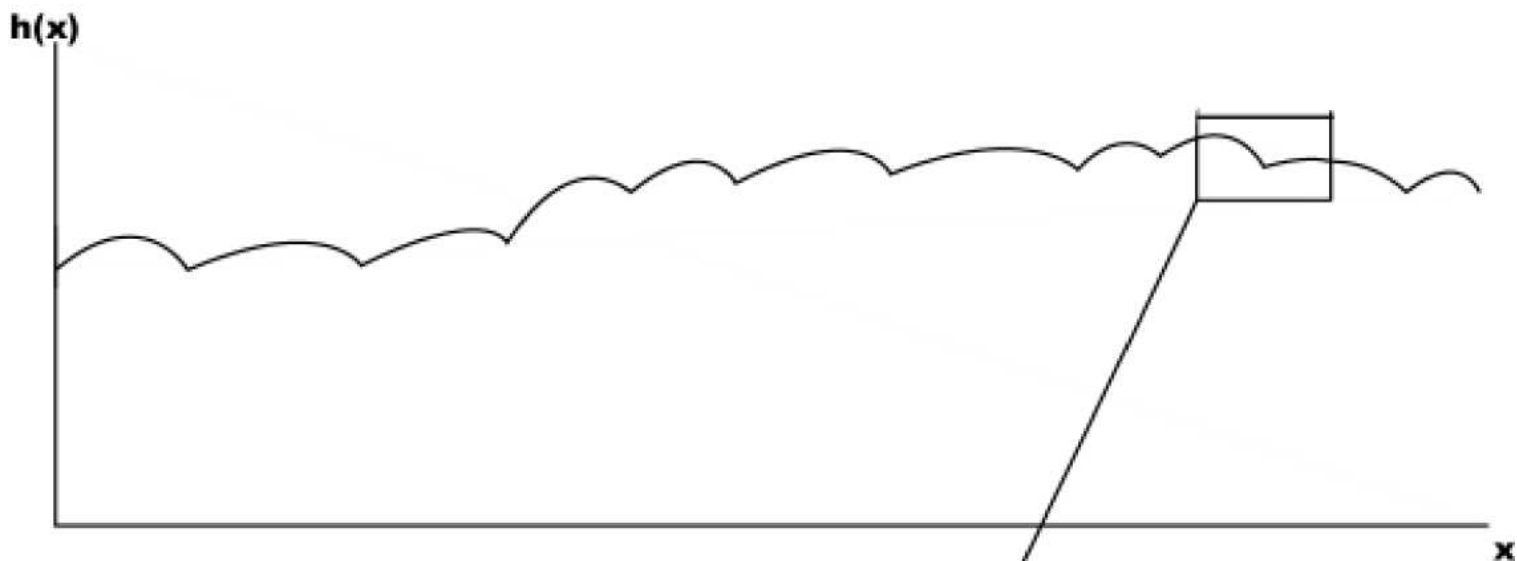
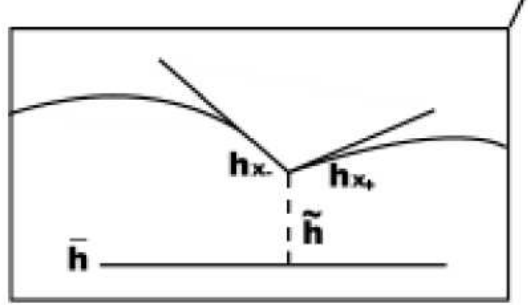
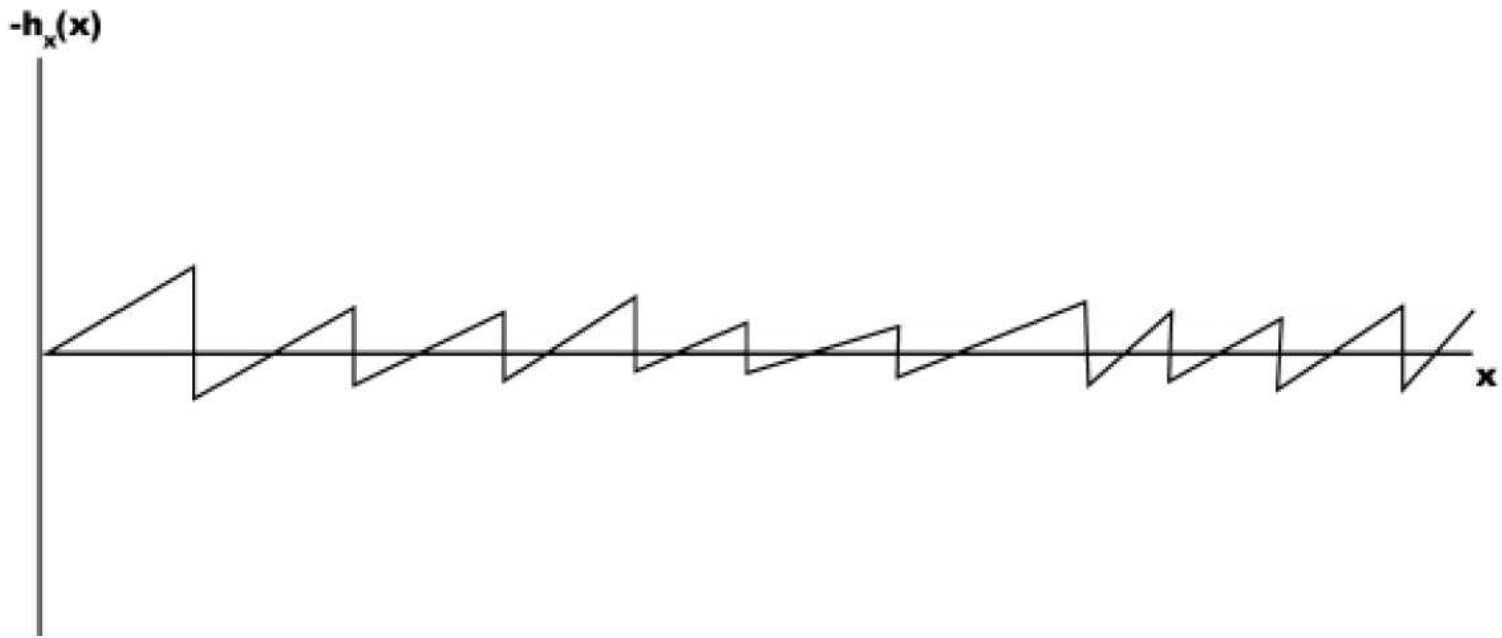
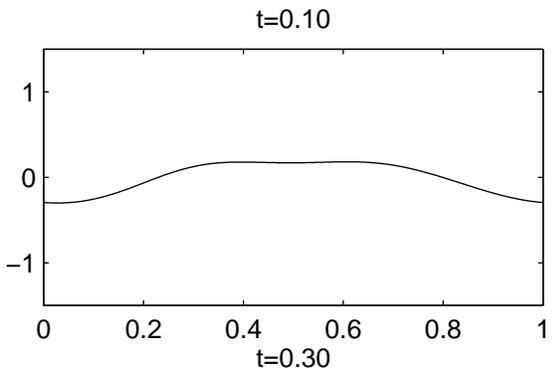
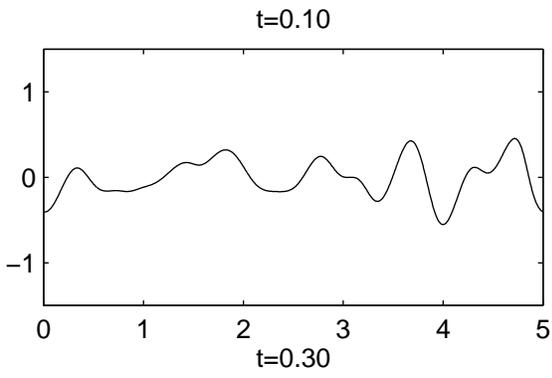
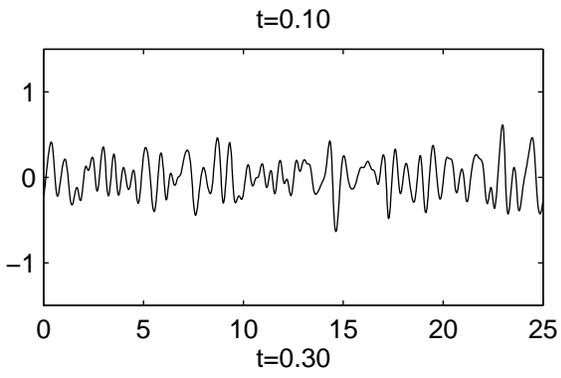
\end{document}